\documentclass[a4paper,fleqn,usenatbib]{mnras}
\pdfoutput=1
\usepackage{amsmath}	% Advanced maths commands
\usepackage{txfonts}

\usepackage[T1]{fontenc}
\usepackage{ae,aecompl}

\usepackage{graphicx}	% Including figure files
\usepackage{amssymb}	% Extra maths symbols

\title[Grackle]{Grackle: a Chemistry and Cooling
  Library for Astrophysics}

\author[B.D. Smith et al.]
       {Britton D. Smith,$^{1,2}$\thanks{E-mail: brittonsmith@gmail.com
           (BDS)},
        Greg L. Bryan$^{3,4}$,
        Simon C. O. Glover$^{5}$,
        Nathan J. Goldbaum$^{6}$, \newauthor
        Matthew J. Turk$^{7,8}$,
        John Regan$^{9,10}$,
        John H. Wise$^{11}$,
        Hsi-Yu Schive$^{6}$,
        Tom Abel$^{12,13}$, \newauthor
        Andrew Emerick$^{3,14}$,
        Brian W. O'Shea$^{15,16,17}$,
        Peter Anninos$^{18}$, \newauthor
        Cameron B. Hummels$^{19}$,
        and Sadegh Khochfar$^{1}$\\
$^{1}$Institute for Astronomy, University of Edinburgh, Royal
Observatory, Edinburgh EH9 3HJ, UK\\
$^{2}$San Diego Supercomputer Center, University of California, San Diego,
10100 Hopkins Drive, La Jolla, CA 92093\\
$^{3}$Columbia University, Department of Astronomy, New York, NY,
10025, USA\\
$^{4}$Simons Center for Computational Astrophysics, New York, NY,
USA\\
$^{5}$Universit\"{a}t Heidelberg, Zentrum f\"{u}r Astronomie, Institut
f\"{u}r Theoretische Astrophysik, Albert-Ueberle-Stra{\ss}e 2, \\69120
Heidelberg, Germany\\
$^{6}$National Center for Supercomputing Applications, University of
Illinois, Urbana-Champaign, IL, 61820, USA\\
$^{7}$School of Information Sciences, University of Illinois,
Urbana-Champaign, IL, 61820, USA\\
$^{8}$Department of Astronomy, University of Illinois,
Urbana-Champaign, IL, 61820, USA\\
$^{9}$Institute for Computational Cosmology, Durham University, South
Road, Durham, DH1 3LE, UK \\
$^{10}$School of Mathematical Sciences, Dublin City University, Dublin,
Ireland \\
$^{11}$Center for Relativistic Astrophysics, Georgia Institute of
Technology, 837 State Street, Atlanta, GA 30332, USA\\
$^{12}$Kavli Institute for Particle Astrophysics and Cosmology,
Stanford University, Menlo Park, CA 94025, USA\\
$^{13}$Department of Physics, Stanford University, Stanford, CA 94305,
USA\\
$^{14}$American Museum of Natural History, Department of Astrophysics,
New York, NY, USA\\
$^{15}$Department of Physics and Astronomy, Michigan State University,
East Lansing, MI 48824, USA\\
$^{16}$Department of Computational Mathematics, Science and
Engineering, Michigan State University, East Lansing, MI 48824, USA\\
$^{17}$JINA: Joint Institute for Nuclear Astrophysics, Michigan State
University, East Lansing, MI 48824, USA\\
$^{18}$Lawrence Livermore National Laboratory, Livermore, CA 94550\\
$^{19}$California Institute of Technology, Pasadena, CA 91125, USA\\
}

\date{Accepted XXX. Received YYY; in original form ZZZ}

\pubyear{2016}

\begin{document}
\label{firstpage}
\pagerange{\pageref{firstpage}--\pageref{lastpage}}
\maketitle

% Abstract of the paper
\begin{abstract}
We present the \texttt{Grackle} chemistry and cooling library for
astrophysical simulations and models.  \texttt{Grackle} provides
a treatment of non-equilibrium primordial chemistry and cooling for H, D, and He
species, including H$_{2}$ formation on dust grains; tabulated
primordial and metal cooling; multiple UV background models; and
support for radiation transfer and arbitrary heat sources.  The
library has an easily implementable interface for simulation codes
written in C, C++, and Fortran as well as a Python interface with
added convenience functions for semi-analytical models.  As an
open-source project, \texttt{Grackle} provides a community resource
for accessing and disseminating astrochemical data and numerical
methods.  We present the full details of the core functionality, the
simulation and Python interfaces, testing infrastructure, performance,
and range of applicability.  \texttt{Grackle} is a fully open-source
project and new contributions are welcome.
\end{abstract}

% Select between one and six entries from the list of approved keywords.
% Don't make up new ones.

%%%%%%%%%%%%%%%%%%%%%%%%%%%%%%%%%%%%%%%%%%%%%%%%%%

%%%%%%%%%%%%%%%%% BODY OF PAPER %%%%%%%%%%%%%%%%%%

\section{Introduction} \label{sec:intro}

\subsection{Why are chemistry and radiative cooling necessary in
  simulations?}

Modeling of plasma chemistry and radiative cooling is absolutely
essential in a wide range of astrophysical phenomena.  At a
fundamental level, virtually all astrophysical objects begin as a
cloud of diffuse plasma in a gravitational potential that is created
primarily either by the plasma itself (i.e., stars) or by a dark
matter halo (i.e., cosmological structure).  In these situations, in
the absense of any additional physical processes, the plasma will
arrange itself so that pressure is in rough equilibrium with gravity,
and no further evolution will occur without some external influence.
Some physical process that allows this plasma to lose energy is
necessary in order to break this stalemate, thus allowing the
formation of stars and galaxies.  The process that typically enables
this energy loss is radiative cooling, often facilitated by a series
of chemical reactions that further enhance the plasma's ability to
lose energy.

Radiative cooling plays a critical role in several important
astrophysical processes.  There is a complex interplay of gas- and
dust-phase chemistry in star formation, which completely dominates the
dynamics of the evolving pre-stellar cloud, and may strongly influence
the resulting stellar initial mass function
\citep{2002Sci...295...93A, 2011ApJ...726...55T,
2008AIPC..990...25G}. At 
a smaller physical scale, radiative cooling can profoundly affect the
structure and behavior of accretion disks around stars and compact
objects \citep{1973A&A....24..337S}. The shape of the `cooling curve'
of diffuse astrophysical plasmas -- i.e., the cooling rate as a
function of density and temperature -- is responsible for the thermal
instability that generates a multiphase interstellar medium
\citep{1977ApJ...218..148M, 1978ppim.book.....S}.  In cosmological
structure formation, gas collapsing into dark matter halos generally
experiences a strong shock at roughly the virial radius, which heats
the gas to roughly the virial temperature.  Optically-thin radiative
cooling of this plasma allows gas to concentrate at the center of dark
matter halos, and ultimately to form molecular clouds and stars
\citep{1977MNRAS.179..541R, 1991ApJ...379...52W}.

Chemistry is often an important component of the evolution of
astrophysical plasmas.  The creation of simple molecules via gas- and
dust-phase chemical reactions can greatly enhance the efficacy of
cooling \citep{1979ApJS...41..555H, 2005ApJ...626..627O}.  The formation
and destruction of simple molecules can be energetically important in
some circumstances, such as Population III star formation
\citep{1998ApJ...508..141O, 2002Sci...295...93A, 2008MNRAS.388.1627G,
2009Sci...325..601T}.
And, from a dynamical perspective, the non-equilibrium evolution of
particular ions can have a strong effect on the ability of gas to cool
\citep{1997NewA....2..181A, 1997NewA....2..209A} and on common
observational quantities such as the column density of O {\sc vi} absorption
line systems in the intergalactic medium that are used to estimate the
metal content of the intergalactic medium and to trace the ``missing
baryons'' \citep{2006ApJ...650..573C, 2011ApJ...731....6S,
  2013MNRAS.430.1548H, 2013MNRAS.434.1043O, 2014ApJ...796...49S}.

\subsection{Why is a multi-code library a good thing?}

Utilizing identical implementations of microphysical solvers, such as
those for chemistry, provides several distinct advantages to the
developers and users of simulation codes as well as the broader
community of researchers who synthesize results of those simulation
codes.

Firstly, and most importantly, the development of a microphysical solver that
can be applied to multiple independent simulation codes reduces the technical,
social and cognitive overhead for collaboration amongst simulators.
This collaboration is a positive benefit in and of itself (particularly as it
reduces duplication of effort) but also provides opportunities to share
understanding, propagate bug fixes, and also to collaboratively implement 
algorithms that provide higher-fidelity results.  Increasing the ability of
researchers to directly collaborate around technology will increase overall
productivity.

With libraries such as the one presented here, where there is a set of
reference
physical values (such as chemical kinetic rate coefficients, drawn from
experimentation or detailed theoretical calculation), the existence of such a
library provides a fixed reference point for calculations that utilize those
reference values.  For instance, by citing a particular version of
the libary in a simulation paper, researchers can indicate which set of
values were used in the calculation.  In cases where results are found to
sensitively depend on these values, or when these reference values are updated,
this can guide deeper physical understanding.

An often underrecognized benefit of portable libraries
is the impact they have on the education and training of early-stage researchers.
For researchers at early stages in their career (such as graduate students and
postdocs) where changing institutions may mean utilizing a different simulation
platform, a shared chemistry package enables them to immediately transfer
developed code to new projects, rather than having to reimplement in a new
system.  Additionally, exposure to such a code that is a mix
of legacy and modern development styles, from several different maintainers,
provides insight into practices in computational science.

Perhaps the most obvious benefit to utilizing a multi-platform library such as
that which we shall now present, particularly in domains such as galaxy formation (where
controlled comparisons are necessary to decouple effects of star formation
prescriptions), is the simplification of directly comparing and reducing the
number of potential sources of difference between multiple calculations.

\subsection{Introducing Grackle}

In this paper, we introduce the \texttt{Grackle} chemistry and cooling
library for astrophysical simulations.  The aim of the
\texttt{Grackle} project is to provide all of the benefits outlined
above: to provide a community resource for accessing and disseminating
data and methods, create a citable version history for evolving
functionality, simplify comparison of results, and reduce the overhead
for collaboration.  In Section \ref{sec:origins}, we give an account
of the history of the \texttt{Grackle} source code and its development
leading up to this publication.  Following this, we describe the
library in full.  In Section \ref{sec:primordial_chemistry}, we detail
the non-equilibrium primordial chemistry solver.  In Section
\ref{Cooling}, we discuss the radiative cooling and heating processes
that are included.  In Section \ref{section:radback}, we describe
the treatment of UV radiation backgrounds.  In Section
\ref{methods:code}, we discuss implementation details, including the
simulation code API, the Python interface, the organization of
the source code, and the procedure for adding to the code.  In Section
\ref{sec:profiling-and-testing}, we describe the testing
infrastructure, the optimization strategy, and present performance
metrics.  Finally, we conclude in Section \ref{sec:summary} with a
discussion of the physical conditions in which the code is valid,
other limitations that should be considered, a list of simulation
codes known to have implemented \texttt{Grackle}, and some future
directions for the project.  Table \ref{tab:resources} lists the
locations of important \texttt{Grackle}-related resources.

\begin{table}
  \centering
  \caption{Important \texttt{Grackle} resources.}
  \label{tab:resources}
  \begin{tabular}{ll}
    \hline
    Source code & https://bitbucket.org/grackle/grackle\\
    Documentation & https://grackle.readthedocs.io/\\
    Mailing list & ``grackle-cooling-users'' on Google Groups\\
    \hline
  \end{tabular}
\end{table}

\section{Origins and Early History}
\label{sec:origins}

The original \texttt{Grackle} source code dates back to work done in
1995 by Peter Anninos and collaborators
\citep{1997NewA....2..209A} who developed a static Eulerian code
focused on primordial gas (e.g. Lyman alpha forest, first stars).
This code was then incorporated into the \texttt{Enzo} codebase in 2000
and modified to include more rates and physical processes.  Metal
cooling using tables from \texttt{Cloudy} was added in 2007
\citep{2008MNRAS.385.1443S}.  In 2012, the AGORA simulation comparison
project \citep{2014ApJS..210...14K} was first organized with the goal
of using a ``common physics package'' to eliminate differences in
results due to the use of cooling solvers that were specific to each
simulation code.  The modularity of \texttt{Enzo}'s cooling solver made
it relatively straightforward to extract it from the \texttt{Enzo}
source.  The initial commit to the \texttt{Grackle} repository,
marking the extraction of the core chemistry and cooling machinery,
was made on October 2, 2012.  The first official release
(\texttt{Grackle} 1.0) occurred on January 10, 2014, with five
additional major releases following over the next two and half years.
Integer increments of the version number marked changes to the
application program interface (API), while decimal releases included
only new features and bugfixes.  A list of all major releases is given
in Table \ref{tab:releases}.  A summary of \texttt{Grackle}'s
development history can be found at
\url{https://www.openhub.net/p/thegrackle}.

\begin{table}
  \centering
  \caption{Dates of major \texttt{Grackle} releases.}
  \label{tab:releases}
  \begin{tabular}{ll}
    \hline
    Version & Release Date\\
    \hline
    1.0 & January 10, 2014\\
    1.1 & October 1, 2014\\
    2.0 & October 1, 2014\\
    2.1 & June 3, 2015\\
    2.2 & May 18, 2016\\
    3.0 & November 1, 2016\\
    \hline
  \end{tabular}
\end{table}

%\section{Chemistry and Cooling Methods} \label{sec:physics}

% ========== Primoridal Chemistry =========

\section{Primordial Chemistry} \label{sec:primordial_chemistry}
The treatment of primordial chemistry (i.e.\ the chemistry of metal-free gas) used in \texttt{Grackle} is closely based on the
treatment in the Enzo AMR code \citep{2014ApJS..211...19B}, although \texttt{Grackle} accounts for a few processes that are not included 
in the latest version of Enzo available at the time of writing (version 2.5). The Enzo primordial chemistry itself is based 
originally on the work of \citet{1997NewA....2..181A} and \citet{1997NewA....2..209A}, although the current version has been
modified substantially compared to the original Abel~et~al.\ network. In this section, we describe in detail the chemistry 
included in \texttt{Grackle} and discuss how the resulting set of chemical rate equations is solved. 

\subsection{Chemistry Network} \label{sec:network}
\texttt{Grackle} provides three different primordial chemistry networks, differing in the number of chemical species that they include. 
The choice of chemical network is controlled by the \texttt{primordial\_chemistry} parameter. Setting \texttt{primordial\_chemistry = 1} selects the
six species network, which tracks the abundances of the species H, H$^{+}$, He, He$^{+}$, He$^{++}$ and e$^{-}$ and is 
designed for modelling atomic and/or ionized gas. Setting \texttt{primordial\_chemistry = 2} selects the nine species network. This
includes all of the species and reactions included in the six species network, but adds molecular hydrogen (H$_2$), plus the
two ions primarily responsible for its formation in primordial gas (H$^{-}$ and H$_{2}^{+}$). Finally, setting \texttt{primordial\_chemistry = 3}
selects the twelve species network, which is an extension of the nine species model that includes D, D$^{+}$ and HD.

\begin{table}
\caption{Chemical reactions in the six species network \label{tab:six}}
\begin{tabular}{lclcc}
\hline
\multicolumn{3}{c}{Reaction} & \multicolumn{2}{c}{Reference} \\
& & & Data & Fit \\
\hline
${\rm H + e^{-}}$ & $\rightarrow$ & $\rm{H^{+} + e^{-} + e^{-}}$ & 1 & 2 \\
${\rm H^{+} + e^{-}}$ & $\rightarrow$ & $\rm{H + \gamma} $ & 3 & 2, 4\\
${\rm He + e^{-}}$ & $\rightarrow$ & ${\rm He^{+} + e^{-} + e^{-}}$ & 1 & 2 \\
${\rm He^{+} + e^{-}}$ & $\rightarrow$ & ${\rm He + \gamma}$ & 5, 6 & 4, 6, 7 \\
${\rm He^{+} + e^{-}}$ & $\rightarrow$ & ${\rm He^{++} + e^{-} + e^{-}}$ & 1 & 2 \\
${\rm He^{++} + e^{-}}$ & $\rightarrow$ & ${\rm He^{+} + \gamma}$ & 3, 8 & 4, 9 \\
${\rm H + H}$ & $\rightarrow$ & ${\rm H^{+} + e^{-} + H}$ & 10 & 11 \\
${\rm H + He}$ & $\rightarrow$ & ${\rm H^{+} + e^{-} + He}$ & 12 & 11 \\
${\rm H + \gamma}$ & $\rightarrow$ & ${\rm H^{+} + e^{-}}$ & 13 & --- \\
${\rm He + \gamma}$ & $\rightarrow$ & ${\rm He^{+} + e^{-}}$ & 13 & --- \\
${\rm He^{+} + \gamma}$ & $\rightarrow$ & ${\rm He^{++} + e^{-}}$ & 13 & --- \\
\hline
\end{tabular}
\\ Key: 1 -- \citet{1987ephh.book.....J}; 2 -- \citet{1997NewA....2..181A}; 3 -- \citet{1992ApJ...387...95F}; 4 -- \citet{1997MNRAS.292...27H}; 5 -- \citet{1960MNRAS.121..471B}; 6 -- \citet{1973A&A....25..137A}; 7 -- \citet{1981MNRAS.197..553B}; 8 -- \citet{1978ppim.book.....S}; 9 -- \citet{1992ApJS...78..341C}; 10 -- \citet{1987PhRvA..36.3100G}; 11 -- \citet{1991ApJS...76..759L}; 12 -- \citet{1981JChPh..74..314V}; 13 -- see Section~\ref{section:radback}
\end{table}

The reactions included in the six species network are listed in Table~\ref{tab:six}. The rate coefficients for these reactions are implemented in
\texttt{Grackle} using simple temperature-dependent analytical fits. In the Table, we list the references from which we take these fits, and also the 
references that are the original sources of the theoretical or experimental  data on which these fits are based.

A few of the reactions listed in Table~\ref{tab:six} deserve further comment:
\begin{enumerate}
\item[(i)] By default, the recombination of
H$^{+}$, He$^{+}$ and He$^{++}$ is modelled using the case A
recombination rate coefficients (the optically-thin approximation in
which recombination photons above 1 Ryd escape). However, the case B
rate coefficients \citep[in which recombination photons above 1 Ryd are
locally re-absorbed,][]{1989agna.book.....O} can instead be selected by setting \texttt{CaseBRecombination} = 1. The additional complication that photons from the recombination of
He$^{+}$ can bring about the photoionization of hydrogen \citep[discussed in some detail in][]{1989agna.book.....O} is not accounted
for, but in most circumstances this will only lead to a small error in the H$^{+}$ and He$^{+}$ abundances. 
\item[(ii)] The rates of the photoionization reactions are not calculated internally by \texttt{Grackle}, but instead are specified either via an input data file or 
passed directly to \texttt{Grackle}  via the  \texttt{Grackle} API as described in Section~\ref{section:radback}.
\item[(iii)] The six species network implemented in \texttt{Grackle} includes two additional reactions that were not part of the original \citet{1997NewA....2..181A}
six species network: the collisional ionization of H by collisions with H and He atoms. Often, these reactions are unimportant. However, they 
can become competitive with the collisional ionization of H by electrons if the fractional ionization of the gas is very low \citep[see, e.g.][for an example of when this can be important]{2015MNRAS.451.2082G}.
\end{enumerate}

\begin{table}
\caption{Chemical reactions in the nine species network \label{tab:nine}}
\begin{tabular}{lclcc}
\hline
\multicolumn{3}{c}{Reaction} & \multicolumn{2}{c}{Reference} \\
& & & Data & Fit \\
\hline
${\rm H + e^{-}}$ & $\rightarrow$ & ${\rm H^{-} + \gamma}$ & 1 & 2 \\
${\rm H^{-} + H}$ & $\rightarrow$ & ${\rm H_{2} + e^{-}}$ & 3 & 3 \\
${\rm H + H^{+}}$ & $\rightarrow$ & ${\rm H_{2}^{+} + \gamma}$ & 4 & 5 \\
${\rm H_{2}^{+} + H}$ & $\rightarrow$ & ${\rm H_{2} + H^{+}}$ & 6 & 6 \\
${\rm H_{2} + H^{+}}$ & $\rightarrow$ & ${\rm H_{2}^{+} + H}$ & 7 & 8 \\
${\rm H_{2} + e^{-}}$ & $\rightarrow$ & ${\rm H + H + e^{-}}$ & 9 & 9 \\
${\rm H_{2} + H}$ & $\rightarrow$ & ${\rm H + H + H}$ & 10 & 10 \\
${\rm H^{-} + e^{-}}$ & $\rightarrow$ & ${\rm H + e^{-} + e^{-}}$ & 11 & 12 \\
${\rm H^{-} + H}$ & $\rightarrow$ & ${\rm H + e^{-} + H}$ & 11 & 12 \\
${\rm H^{-} + H^{+}}$ & $\rightarrow$ & ${\rm H + H}$ & 13 & 14 \\
${\rm H^{-} + H^{+}}$ & $\rightarrow$ & ${\rm H_{2}^{+} + e^{-}}$ & 15 & 12, 16 \\
${\rm H_{2}^{+} + e^{-}}$ & $\rightarrow$ & ${\rm H + H}$ & 17 & 12 \\
${\rm H_{2}^{+} + H^{-}}$ & $\rightarrow$ & ${\rm H_{2} + H}$ & 18 & 18 \\ 
${\rm H + H + H}$ & $\rightarrow$ & ${\rm H_{2} + H}$ & 19  & 19  \\
${\rm H + H + H_{2}}$ & $\rightarrow$ & ${\rm H_{2} + H_{2}}$ & 20, 21 & 22 \\
${\rm H^{-} + \gamma}$ & $\rightarrow$ & ${\rm H + e^{-}}$ & 23 & --- \\
${\rm H_{2}^{+} + \gamma}$ & $\rightarrow$ & ${\rm H + H^{+}}$ & 23 & --- \\
${\rm H_{2} + \gamma}$ & $\rightarrow$ & ${\rm H_{2}^{+} + e^{-}}$ & 23 & --- \\
${\rm H_{2}^{+} + \gamma}$ & $\rightarrow$ & ${\rm H^{+} + H^{+} + e^{-}}$ & 23 & --- \\
${\rm H_{2} + \gamma}$ & $\rightarrow$ & ${\rm H + H}$ & 23 & --- \\
${\rm H + H + grain}$ & $\rightarrow$ & ${\rm H_{2} + grain}$ & --- & 24$^{*}$ \\
\hline
\end{tabular}
\\ Note: the nine species network also includes all of the reactions listed in Table~\ref{tab:six}.
\\ Key: 1 -- \citet{1979MNRAS.187P..59W}; 2 --
\citet{1998ApJ...509....1S}; 3 -- \citet{2010Sci...329...69K}; 4 --
\citet{1976PhRvA..13...58R}; 5 -- \citet{2015MNRAS.446.3163L}; 6 --
\citet{1979JChPh..70.2877K}; 7 -- \citet{2002PhRvA..66d2717K}; 8 ---
\citet{2004ApJ...606L.167S,2004ApJ...607L.147S}; 9 --
\citet{2002PPCF...44.1263T}; 10 -- \citet{1996ApJ...461..265M}; 11 --
\citet{1987ephh.book.....J}; 12 -- \citet{1997NewA....2..181A}; 13 --
\citet{1986JPhB...19L..31F}; 14 -- \citet{1999MNRAS.304..327C}; 15 --
\citet{1978JPhB...11L.671P}; 16 -- \citet{1987ApJ...318...32S}; 17 --
\citet{1994ApJ...424..983S}; 18 -- \citet{1987IAUS..120..109D}; 19 --
See text; 20 -- \citet{1962JChPh..36.2923S}; 21 --
\citet{1970JChPh..53.4395H}; 22 -- \citet{1983JPCRD..12..531C}; 23 --
see Section~\ref{section:radback}; 24 -- \citet{1985ApJ...291..722T,
  2000ApJ...534..809O}; * - This reaction included as an additional
option when metals are present.
\end{table}

\begin{figure}
  \centering
  \includegraphics[width=0.45\textwidth]{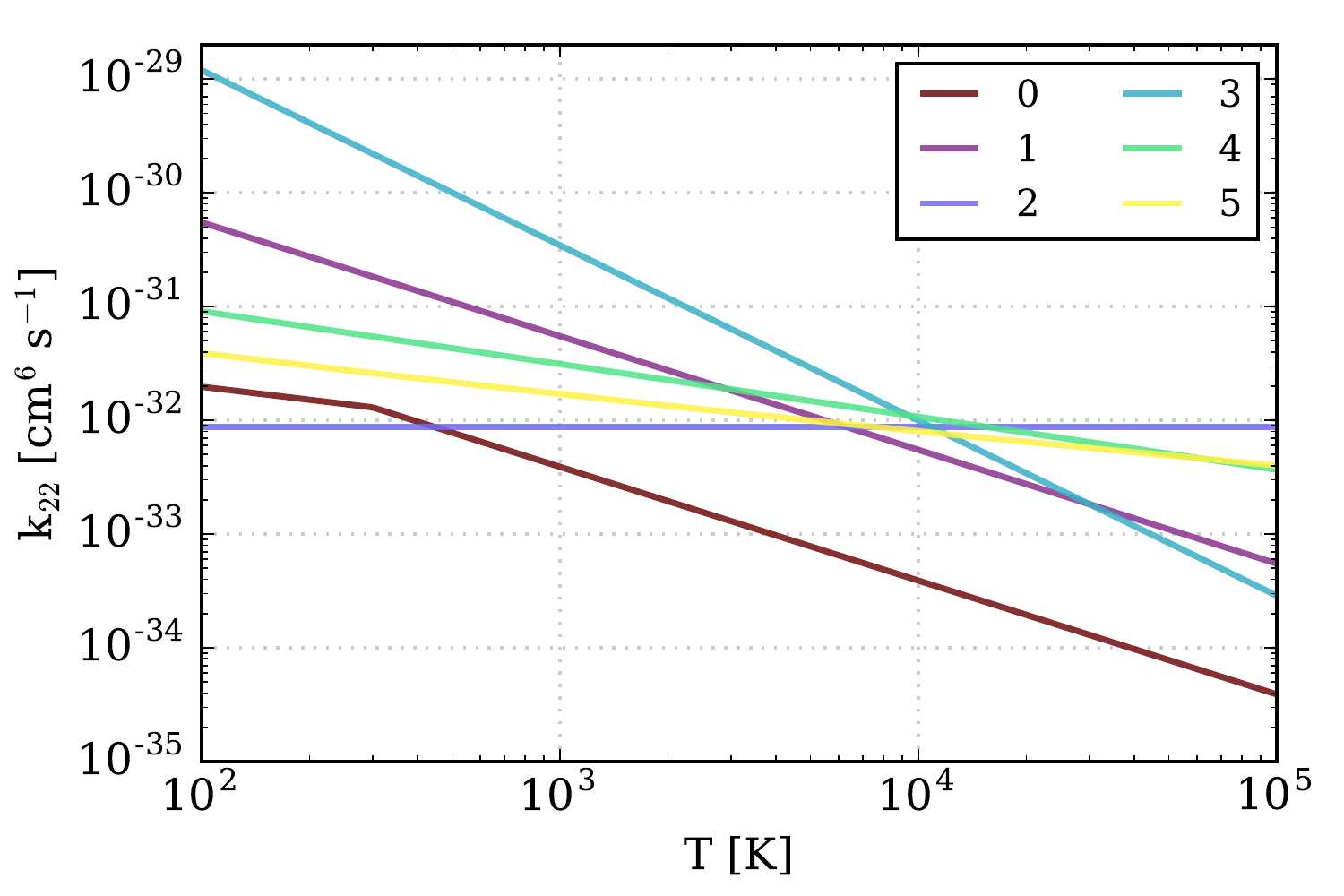}
  \caption{ The available three-body H$_{2}$ formation rates as a
    function of temperature: 0 - \citet{2002Sci...295...93A}, 1 -
    \citet{1983ApJ...271..632P}, 2 - \citet{1983JPCRD..12..531C}, 3 -
    \citet{2007MNRAS.377..705F}, 4 - \citet{2008AIPC..990...25G}, 5 -
    \citet{2013ApJ...773L..25F}.} \label{fig:threebody}
\end{figure}

The nine species network includes all of the reactions in Table~\ref{tab:six} plus the additional reactions listed in Table~\ref{tab:nine}. Again,
a couple of reactions deserve further discussion:
\begin{enumerate}
\item[(i)] The treatment of H$_{2}$ collisional dissociation by H atom collisions now used in \texttt{Grackle} is taken from \citet{1996ApJ...461..265M} and 
accounts for both the temperature and the density dependence of this process. It therefore remains valid in the high density limit, where the H$_{2}$
level populations approach their local thermodynamical equilibrium (LTE) values. This is important, because H$_{2}$ is far more susceptible to
collisional dissociation in this limit than when it is solely in the vibrational ground state. It should also be noted that the treatment of this process
in \texttt{Grackle} accounts for effects of dissociative tunneling as well as direct collisional dissociation; previously, Enzo only accounted for the latter
process and hence underestimated the dissociation rate at low temperatures \citep{2014MNRAS.443.1979L,2015MNRAS.451.2082G}
\item[(ii)] In view of the considerable uncertainty in the rate of the three-body reaction
\begin{equation}
{\rm H + H + H} \rightarrow {\rm H_{2} + H},
\end{equation}
discussed in detail in \citet{2008AIPC..990...25G} and \citet{2011ApJ...726...55T}, \texttt{Grackle} provides several different rate coefficients for this process. The user can
select which of these rate coefficients to adopt by means of the \texttt{three\_body\_rate} parameter. The options are
\begin{enumerate}
\item[{\bf 0}:]  Rate coefficient from \citet{2002Sci...295...93A}, based on an extrapolation from low temperature calculations by \citet{1987JChPh..87..314O}. This is the default option.
\item[{\bf 1}:]  Rate coefficient from \citet{1983ApJ...271..632P}, derived using detailed balance applied to the H$_{2}$ collisional dissociation rate measured by \citet{1967JChPh..47...54J}.
\item[{\bf 2}:]  Rate coefficient recommended by \citet{1983JPCRD..12..531C}, based on a survey of the experimental data available at that time. 
\item[{\bf 3}:]  Rate coefficient from \citet{2007MNRAS.377..705F}, also derived from \citet{1967JChPh..47...54J} using detailed balance, but with a different treatment of the H$_{2}$ partition
function. 
\item[{\bf 4}:]  Rate coefficient from \citet{2008AIPC..990...25G}, derived from the \citet{1996ApJ...461..265M} high-density H$_2$ collisional dissociation rate using detailed balance
\item[{\bf 5}:]  Rate coefficient computed directly by \citet{2013ApJ...773L..25F}.
\end{enumerate}
\end{enumerate}
We plot each of these rates in Figure \ref{fig:threebody}.

Finally, the twelve species network includes the reactions in Tables~\ref{tab:six} and \ref{tab:nine}, plus a small number of additional reactions involving
D$^{+}$, D and HD, listed in Table~\ref{tab:12}. The intent of the twelve species network is to allow the HD abundance of the gas to be tracked accurately,
since in cold gas HD can become a more effective coolant than H$_{2}$ despite its much lower fractional abundance \citep[see e.g.][]{2006MNRAS.366..247J,2008ApJ...685....8M}. It is therefore necessary to include only a small number of reactions, as the direct conversion of H$_{2}$ to HD by collisions with D$^{+}$ and D, together with the
corresponding inverse reactions generally dominate the production and destruction of HD.

Two reactions in the twelve species network require further discussion:
\begin{enumerate}
\item[(i)] The rate coefficient that we adopt for the reaction
\begin{equation}
{\rm HD + H} \rightarrow {\rm H_{2} + D}
\end{equation}
is an analytical fit presented in \citet{2002P&SS...50.1197G}, based on data from \citet{1959JChPh..31.1359S}. However, this fit blows up
at temperatures $T < 100$~K, yielding an unphysically large value for the rate coefficient. We therefore follow \citet{2007MNRAS.376..709R}
and \citet{2008ApJ...685....8M} and assume that the rate at $T < 100$~K is the same as the rate at $T = 100$~K. Note that as this reaction
proceeds extremely slowly at temperatures below a few hundred K, this is unlikely to be a significant source of error. 
\item[(ii)] We assume that the rate coefficient for the associative detachment of H$^{-}$ by D
\begin{equation}
{\rm D + H^{-} \rightarrow  HD + e^{-}}
\end{equation}
is the same as for the corresponding reaction between H and H$^{-}$, since measurements by \citet{2012PhRvA..86c2714M} suggest that
there is not a significant isotope effect for this reaction. However, in the solver, we multiply the rate coefficient by a factor of two when
computing the HD formation rate to account approximately for the contribution from the reaction
\begin{equation}
{\rm H + D^{-} \rightarrow  HD + e^{-}}.
\end{equation}
Note that we do not explicitly include this reaction in our network because it would require us to track the abundance of the D$^{-}$ ion,
thereby adding significant additional complexity to the model for only a marginal increase in accuracy.
\end{enumerate}

\begin{table}
\caption{Additional reactions included in the twelve species network \label{tab:12}}
\begin{tabular}{lclcc}
\hline
\multicolumn{3}{c}{Reaction} & \multicolumn{2}{c}{Reference} \\
& & & Data & Fit \\
\hline
${\rm H^{+} + D}$ & $\rightarrow$ & ${\rm H + D^{+}}$ & 1, 2 & 3 \\
${\rm D^{+} + H}$ & $\rightarrow$ & ${\rm D + H^{+}}$ & 1, 2 & 3 \\
${\rm H_{2} + D^{+}}$ & $\rightarrow$ & ${\rm HD + H^{+}}$ & 4 & 5 \\
${\rm HD + H^{+}}$ & $\rightarrow$ & ${\rm H_{2} + D^{+}}$ & 4 & 5 \\
${\rm H_{2} + D}$ & $\rightarrow$ & ${\rm HD + H}$ & 6 & 7 \\
${\rm HD + H}$ & $\rightarrow$ & ${\rm H_{2} + D}$ & 8 & 5, 9 \\
${\rm D + H^{-}}$ & $\rightarrow$ & ${\rm HD + e^{-}}$ & 10, 11 & 10  \\
\hline
\end{tabular}
\\ Note: the twelve species network also includes all of the reactions listed in Tables~\ref{tab:six} \& \ref{tab:nine}.
\\ Key: 1 -- \citet{1999PhRvL..83.4041I}; 2 -- \citet{2000PhRvA..62d2706Z}; 3 -- \citet{2002ApJ...566..599S};
4 -- \citet{1982sasp.nasa..304G}; 5 -- \citet{2002P&SS...50.1197G}; 6 -- \citet{2003PhRvL..91f3201M}; 7 -- \citet{2011ApJ...727..110C}; 
8 -- \citet{1959JChPh..31.1359S}; 9 -- \citet{2007MNRAS.376..709R}; 10 -- \citet{2010Sci...329...69K}; 11 -- \citet{2012PhRvA..86c2714M}
\end{table}

% ----------------------------------

\subsection{Solving and Updating the Network}

Chemical networks such as the ones described above are often challenging to evolve due to the very different time scales that various rates may have -- creation and destruction time scales can differ by orders of magnitude among the different species.   Such ``stiff" sets of equations are often solved with implicit methods, which permit longer timesteps.  Several packages exist to do this, and can even switch between methods for solving stiff and non-stiff equations, such as LSODAR \citep{Hindmarsh83}; however, for multi-dimensional simulations it is useful to have an implementation which is optimized for the case at hand.

Our kinetic network for solving the rate of change of species density $n_i$ has the general form
\begin{equation}
\frac{\partial n_i}{\partial t} = \sum_j \sum_l k_{jl} n_j n_l + \sum_j I_j n_j
\label{eq:rate_general}
\end{equation}
where $k_{jl}$ is the rate for reactions involving species $j$ and $l$
while $I_j$ is the appropriate radiative rate.  Note, that the
terms on the right-hand side of Equation \ref{eq:rate_general} can
denote either creation or destruction reactions.  For creation
reactions, the corresponding term is positive; for destruction
reactions, it is negative.  In rare cases there may be an additional term for three-body reactions.  To solve this kinetic network, \texttt{Grackle} follows closely the procedure described in \citet{1997NewA....2..209A} and \citet{2014ApJS..211...19B} by grouping creation and destruction rates to rewrite Eq.~\ref{eq:rate_general} as,
\begin{equation}
\frac{\partial n_i}{\partial t} = C_i(T, n_j) - D_i(T, n_j) n_i
\label{eq:rateCD}
\end{equation}
where $C_i$ represents the total creation rate of species $i$ (given the temperature $T$ and other species densities) while $D_i n_i$ is the destruction rate of the same species (which must be proportional to $n_i$), including both radiative and collisional processes.  Ideally, we would solve these ordinary differential equations using a higher order method; however, here we adopt a very simple low order backwards difference formula (BDF) due to its stability.  \citet{1997NewA....2..209A} explored a variety of higher-order solution techniques but found that this simple BDF scheme was generally more stable and competitive for the level of accuracy required.  In particular, the BDF version of Equation~\ref{eq:rateCD} is:
\begin{equation}
n^{t + \Delta t} = \frac{C^{t+\Delta t} \Delta t + n^t}{1 + D^{t+\Delta t} \Delta t}
\label{eq:rate_BDF}
\end{equation}
Unfortunately, we are not able to fully implement a BDF scheme due to
the difficulty of evaluating the $C_i$ and $D_i$ at the advanced time.
Instead, we attempt to mimic a BDF method through a set of partial
updates combined with sub-cycling.  The partial forwarding updating is
done by solving the various species in a specified order and using
the updated species densities in the following partial step.  The
ordering developed by \citet{1997NewA....2..209A} was based on empirical tests under a wide range of conditions and is as follows for the six species model: H, H$^+$, e$^-$, He, He$^+$, He$^{++}$.  For the nine species model, we then add H$_2$, H$^-$, and H$_2^+$.  The H$_2^+$ timescale is sufficiently short that it can be decoupled and we use instead the equilibrium value:
\begin{equation}
n_{ {\rm H}_2^+} = \frac{k_9 n_{\rm H} n_{\rm H^+} + k_{11} n_{\rm H_2} n_{\rm H^+} + k_{17} n_{\rm H^-} n_{\rm H^+} + k_{29} n_{\rm H_2}}
   {k_{10} n_{\rm H} + k_{18} n_{\rm e} + k_{19} n_{\rm H^-} + k_{28} + k_{30}}
\end{equation}
Finally, for twelve species model, we add D, D$^+$ and HD.  In each case, the updated species of the previous step are used in the next.  

The time step passed to \texttt{Grackle} (often the hydrodynamic timestep of a parent simulation) can be quite large, which may potentially cause large errors in our low-order chemical integrator.  To get around this issue, we sub-cycle the BDF step described above, constraining the chemical timestep such that the H and e$^-$ abundances change by no more than 10\% in any sub-cycle step of length $\Delta t$:
\begin{equation}
\Delta t = 0.1 \min \left( \frac{n_{\rm H}}{\dot{n}_{\rm H}} , \frac{n_{\rm e^-}}{\dot{n}_{\rm e^-}} \right)
\end{equation}
In some cases, particularly close to equilibrium, we slightly modify this.  After more than 50 subcycle steps (if we have not yet integrated a full hydro timestep), we replace the analytically calculated time derivatives in the above expression with numerical time derivatives (i.e. change from the previous sub-cycle step).  This is helpful when we are close to equilibrium and the integrator is taking very small steps and regularly overshooting the equilibrium value.  
%Finally, when the density is high $n_H > 10^8$ cm$^{-3}$, and the net cooling/heating term is positive (see below), we replace use an estimate of the rate of change of H based on the H equilibrium solution.

% ============ Cooling/Heating =============

\section{Heating and Cooling} \label{Cooling}

\texttt{Grackle} can evolve the Lagrangian energy equation, taking into account a wide range of radiative cooling and heating processes:
\begin{equation}
\frac{de}{dt} = - \dot{e}_{\rm cool} + \dot{e}_{\rm heat}
\end{equation}
In this paper, we split our discussion of radiative cooling/heating and chemistry, but in the code, they are solved together, using a simple first-order integrator.  To enhance accuracy, the integrator is sub-cycled with a timestep constraint 
\begin{equation}
\Delta t \leq 0.1 \frac{e}{\dot{e}}.
\end{equation}
If both chemistry and cooling/heating are turned on, these are integrated at the same time, using a timestep which is a minimum of the chemistry and cooling constraints.  In the following sections, we describe the various supported cooling and heating options.  We begin with radiative cooling and heating due solely to hydrogen and helium, and then turn to heavier atomic elements (``metals"), and finally dust. 

% ----------------------------------

\subsection{Primordial Heating and Cooling}

To solve for the effects of primordial (H and He only) heating and cooling, \texttt{Grackle} includes two options: (1) a non-equilibrium solver, the chemistry part of which is described in detail in Section~\ref{sec:primordial_chemistry}, and (2) a tabulated version, which assumes ionization equilibrium to compute the cooling and heating rates due to primordial chemistry.  In both cases, photo-heating from an external radiation source can be important -- this is described in Section~\ref{section:radback}.

\subsubsection{Non-equilibrium} \label{sec:pri-neq}

We include a variety of cooling rates due to transitions of the non-equilibrium species.  We begin with a list appropriate for the six species network (\texttt{primordial\_chemistry = 1}): 

\begin{enumerate}
\item Collisional excitation cooling rates involving the following species: $n_{\rm e} n_{\rm H}$, $n_{\rm e}^2 n_{\rm He^+}$, and $n_{\rm e} n_{\rm He^+}$ \citep{1981MNRAS.197..553B, 1992ApJS...78..341C};
\item Collisional ionization cooling for $n_{\rm e} n_{\rm H}$, $n_{\rm e} n_{\rm He}$, $n_{\rm e} n_{\rm He^+}$ and $n_{\rm e}^2 n_{\rm He^+}$ \citep{1987ApJ...318...32S, 1992ApJS...78..341C, 1997NewA....2..181A};
\item Recombination cooling: $n_{\rm e} n_{\rm H^+}$, $n_{\rm e} n_{\rm He^+}$, $n_{\rm e} n_{\rm He^{++}}$ \citep{1981MNRAS.197..553B, 1992ApJ...387...95F, 1997MNRAS.292...27H};
\item Bremsstrahlung cooling for all ionized species \citep{1981MNRAS.197..553B};
\item Compton cooling/heating off the CMB \citep{1971phco.book.....P}, and 
\item Photoionization heating for H, He and He$^+$, depending on the ionizing radiation field -- see section~\ref{section:radback} for more details.
\end{enumerate}
In addition to the sources referenced above, we note that most of these rates were tabulated in Appendix B of \citet{1997NewA....2..209A}.

For the nine species version (\texttt{primordial\_chemistry = 2}), we add H$_2$ cooling.  Our default cooling rate is as follows.  At high densities, where the level populations are in LTE and hence depend only on temperature, we use the high-density rate from \citet{1998A&A...335..403G}. For low densities, we computed the cooling rate due to collisions with H, H$_2$, He, H$^+$, and e$^-$ as described in section 2.3 of  \citet{2008MNRAS.388.1627G}.  The exception is that for collisions with e$^-$, we use revised rates from \citet{yoon2008cross} and for H$^+$, we adopt rates from \citet{2011PhRvL.107b3201H} and \citet{2012PhRvL.108j9903H}.  For intermediate densities, we use the smooth density-dependent switch from \citet{1998A&A...335..403G}.  In addition to this cooling function, the code can also (depending on a compile time switch) use an older rate from \citet{1983ApJ...270..578L}. At very high densities, \texttt{Grackle} can also account for the decrease in the H$_{2}$ cooling rate that comes about once the H$_{2}$ lines becomes optically thick. This is treated using a simple density-dependent opacity correction term introduced by \citet{2004MNRAS.348.1019R}. This option is enabled by setting \texttt{h2\_optical\_depth\_approximation = 1}.

\label{sec:chemheat}

In addition to H$_2$ radiative cooling, we include the impact of chemical heating or cooling due to the formation or destruction of molecular hydrogen.  Following \citet{2000ApJ...534..809O}, we add $4.48 (1 + n_{\rm cr}/n)^{-1}$ eV for H$_2$ formation by the three-body reaction, and $0.2 + 4.1(1 + n_{\rm cr}/n)^{-1}$ eV for H$_2$ formation on dust grain surfaces.  The critical density $n_{\rm cr}$ is given by eq. (23) of  \citet{2000ApJ...534..809O}.  For H$_2$ destruction, we remove 4.48 eV per H$_2$ molecule dissociated.

Finally, in twelve species mode (\texttt{primordial\_chemistry = 3}), which adds deuterium chemistry, the code includes (radiative) cooling from HD.  This is a combination of a fit from \citet{2011MNRAS.415..487C} for the high-density limit, and \citet{2007MNRAS.382..133W} for the low density limit.

There are a number of other optional heating and cooling terms that the code includes, some of which are not strictly primordial, but are included as part of this cooling package.  These include:  
\begin{enumerate}
\item Collisionally induced excitation of H$_2$ at high densities, with rates as described in \citet{2004MNRAS.348.1019R}.
\item X-ray Compton heating (or cooling) using eq. (4) and (11) of \citet{1999ApJ...517L...9M}.
\item A photoelectric heating rate, equal to $\Gamma_{\rm eff} n_{\rm H}$, where $\Gamma_{\rm eff}$ is a fixed input parameter.  Although not strictly primordial, we include this rate here as it is distinct from the dust model described later.
\end{enumerate}

\subsubsection{Equilibrium (Tabulated)} \label{sec:pri-tab}

In the other (simpler) mode, the cooling and heating due to the primordial
elements can be calculated using tables of pre-computed values under
the assumption of ionization equilibrium.  If there is
no incident radiation, then we have simple collisional ionization
equilibrium (CIE), and the cooling rate (per hydrogen atom) is solely a function of
temperature.  This means we can look up the cooling rate using a simple
one-dimensional table.   If radiation is present, the cooling rate under
ionization equilibrium for a fixed spectral shape and intensity is a
function of density and temperature, resulting in a two-dimensional table
look-up.  The process by which these tables are
created is discussed in Section \ref{sec:cooling-tables}.  For the
primordial elements, \texttt{Grackle} provides pre-computed tables for the cooling rate, $\Lambda$;
the heating rate, $\Gamma$; and the mean molecular weight, $\mu$, of
the gas as a function of temperature and density, if required.  All rates
are computed using linear interpolation in log-space.

Since simulation codes typically solve for the internal energy of the gas
instead of the temperature, it is necessary to convert one to the
other via
\begin{equation} \label{eqn:e-T}
e = \frac{k T}{(\gamma - 1)\ \mu m_{\rm H}},
\end{equation}
where $k$ is the Boltzmann constant, $T$ is the gas temperature, $e$ is
the specific internal energy, $\gamma$ is the adiabatic index of an
ideal gas, and $m_{\rm H}$ is the mass of a hydrogen atom.  Since
the mean molecular weight, $\mu$, is also a
function of temperature, we solve Equation \ref{eqn:e-T} iteratively
with an initial guess of $\mu = 1$.  The temperature calculated using
the initial guess for $\mu$ is then used as an input to the table of
$\mu(T,...)$, from which a new value of $\mu$ is calculated via
linear interpolation.  To prevent the solution of $\mu$ and $T$ from
oscillating, we apply a dampener such that the new value of $\mu$ is
the average of the old value and the value from the table.  For
iteration, $i$, of this procedure, $\mu_{i}$ is then given by
\begin{equation}
\mu_{i} = \frac{1}{2} (\mu_{i-1} + \mu(T_{i-1},...)).
\end{equation}
To account for the presence of metals, we then apply an additional
correction such that the value with metals included, $\mu_{i, Z}$ is
\begin{equation}
\frac{\rho}{\mu_{i, Z}} = \frac{\rho}{\mu_{i}} +
\frac{\rho_{Z}}{\mu_{Z}},
\end{equation}
where $\rho$ is the total gas density, $\rho_{Z}$ is the metal
density, and $\mu_{Z} \equiv 16$, which
is consistent with a Solar abundance pattern.  In practice, this
process arrives at a solution for $\mu$ and $T$ that converges to
within 1\% in just a few iterations.

Once the gas temperature has been calculated, the cooling and heating
due to the primordial species is then computed via interpolating over
the multidimensional tables.  In the most commonly used mode, heating
comes from a model UV background, which is spatially uniform and
varies as a function of redshift.  In this case, the tables for
$\Lambda$, $\Gamma$, and $\mu$ have dimensions of $z$, $\rho$, and
$T$.  The effects of the UV background models and their implementation
within the code are discussed further in section
\ref{section:radback}.

In a cosmological simulation, the CMB acts as a temperature floor,
below which the gas cannot cool radiatively.  We approximate this effect by
subtracting the cooling rate at the CMB temperature, $T_{\rm CMB}$, from
the calculated cooling rate such that the final cooling rate that is
applied is given by
\begin{equation}
\Lambda_{\rm final}(T) = \Lambda(T) - \Lambda(T_{\rm CMB}).
\end{equation}
This allows the cooling rate to smoothly approach zero as the
temperature approaches $T_{\rm CMB}$ and also for the CMB to heat the gas
when $T < T_{\rm CMB}$.  We take the same approach when calculating the
cooling from metals.

% ----------------------------------

\subsection{Metal Heating and Cooling}

Next we turn to the impact of metals on the thermal evolution of the gas.
Solving for the cooling from metals using a non-equilibrium network
akin to that discussed in section \ref{sec:network} is computationally
challenging since the number of species and reactions that must be
considered rises steeply with each additional element.  For this
reason, \texttt{Grackle} computes the impact from metals using tables of
heating and cooling rates in a way analogous to that discussed in section
\ref{sec:pri-tab}.  This method was first described by
\citet{2008MNRAS.385.1443S}.  Other packages, such as \texttt{KROME}
\citep{2014MNRAS.439.2386G}, offer the ability to perform
non-equilibrium chemistry calculations including metals, but at the
proportional computational cost.

The cooling or heating from metals can be added to the rate from the
primordial species as calculated by either the non-equilibrium network
(\S \ref{sec:pri-neq}) or the tabulated solver (\S
\ref{sec:pri-tab}).  As in the tabulated primordial cooling, the
heating and cooling tables have three dimensions ($z$, $\rho$, and
$T$) to account for the effects of the UV background models.  The
values stored in the tables correspond to those of Solar metallicity
and the cooling rate applied to the gas is scaled by the local
metallicity.  All of the available metal cooling tables assume a Solar
abundance pattern and consider all elements heavier than He up to
atomic number 30 (Zn).

\subsubsection{Constructing Cooling Tables} \label{sec:cooling-tables}

The \texttt{Grackle} library comes with three different model input files that can be
used to calculated the tabulated cooling from primordial species and
metals under different conditions.  The three available models are the
UV background model of \citet{2009ApJ...703.1416F}, that of
\citet{2012ApJ...746..125H}, and a model assuming no incident
radiation, i.e., collisional ionization only.  For the two UV
background models, the input files also contain tables of
photo-ionization, photo-dissociation, photo-detachment, and
photo-heating rates as a function of redshift for various atomic and
molecular H/He species.  These are used in conjunction with the
non-equilibrium primordial chemistry solver.

The cooling tables are
created using the method originally described by
\citet{2008MNRAS.385.1443S}.  Cooling, heating, and mean molecular
weight values are computed using the photoionization simulation code, 
\texttt{Cloudy}\footnote{http://nublado.org/}
\citep{2013RMxAA..49..137F}.  We use the
\texttt{CIALoop}\footnote{https://bitbucket.org/brittonsmith/cloudy\_cooling\_tools}
code of \citet{2008MNRAS.385.1443S} to loop over the appropriate
parameter space, call \texttt{Cloudy}, and collate the results.  To
expedite this process, \texttt{CIALoop} runs in parallel by managing
multiple instances of \texttt{Cloudy} simultaneously.  To calculate
the cooling and heating contribution from metals, we run each of the
above models twice, once with the full complement of elements and once
with only H and He.  For every point in each version of the model, we
extract all cooling/heating components contributing at least
10$^{-10}$ of the total rate.  We then remove all components that
appear in both the full and H/He models, leaving only the
contributions of the metals.  All of the data are organized in
\texttt{HDF5} files.  The structure and discoverability of
\texttt{HDF5} files allows the data to be easily used for other
applications.

% ----------------------------------

\subsection{Dust Heating and Cooling}

Dust grains transfer heat to and from a gas through collisions
with the atoms and molecules in that gas.  The surfaces of dust grains
also provide a site for efficient formation of molecules, particularly
H$_{2}$.  Both heat transfer and molecular formation rates depend very
sensitively on the dust temperature, $T_{\rm gr}$.  The dust temperature
is determined by balancing the relevant heating and cooling terms.
Dust grains are heated by incident radiation and cool through emission
of thermal
radiation.  Heat flows between gas and dust in the direction of
whichever has the lower temperature.  The implementation of
dust-related chemistry here follows very closely the work of
\citet{2000ApJ...534..809O} and \citet{2005ApJ...626..627O}.  As in
these works, we currently assume that heating radiation comes only
from the CMB.  Future versions of the code will allow for the
inclusion of additional radiative heating terms.  This heat balance
equation is, therefore, given by
\begin{equation} \label{eqn:tdust}
4 \sigma T_{\rm gr}^{4} \kappa_{\rm gr} = \Lambda_{\rm gas/grain} + 4 \sigma
T_{\rm rad}^{4} \kappa_{\rm gr},
\end{equation}
where $\sigma$ is the Stefan-Boltzmann constant, $T_{\rm rad}$ is the
radiation temperature, and $\kappa_{\rm gr}$ is the grain opacity.  The
left-hand side of Equation \ref{eqn:tdust} represents cooling by
thermal radiation and the second term on the right-hand side
represents the incident radiation field characterized by a radiation
temperature, $T_{\rm rad}$.  The dust/gas heat transfer rate per unit dust
mass, $\Lambda_{\rm gas/grain}$, is given by
\begin{eqnarray} \label{eqn:gasdust}
\Lambda_{\rm gas/grain} & = & 1.2\times10^{-31}~\frac{n_{\rm H}^{2}}{\rho_{\rm gr}}
\left(\frac{T}{1000 K}\right)^{1/2} (1 - 0.8 e^{-75 / T})  \nonumber \\
& & (T - T_{\rm gr})~\textrm{erg~s$^{-1}$~g$^{-1}$},
\end{eqnarray}
\citep{1989ApJ...342..306H}, where $T$ is the gas temperature and
$\rho_{\rm gr}$ is the dust mass density.  Currently,
the dust to gas mass ratio is assumed to scale with metallicity, i.e.,
the dust to metal mass ratio is constant.  As in
\citet{2000ApJ...534..809O}, we use the grain composition model of
\citet{1994ApJ...421..615P} where the grain mass fraction is
9.34$\times$10$^{-3}$ at Solar metallicity.  In the future, the user
will have the option to provide the dust density independently of the
metal density.  For the grain opacity, we use the piece-wise
polynomial of \citet{2011ApJ...729L...3D}, which is given by
\begin{equation}
\kappa(T_{\rm gr}) \propto \left\{ \begin{array}{ll}
T_{\rm gr}^{2} & \textrm{, $T_{\rm gr}$ $<$ 200~K,}\\
\textrm{constant} & \textrm{, 200~K $<$ $T_{\rm gr}$ $<$ 1500~K,}\\
T_{\rm gr}^{-12} & \textrm{, $T_{\rm gr}$ $>$ 1500~K},
\end{array} \right.
\end{equation}
with a normalization of $\kappa_{\rm gr}(T_{\rm gr} = 200~K) = 16$ cm$^{2}$ g$^{-1}$
\citep{1994ApJ...421..615P, 2000ApJ...534..809O}.  The steep power-law
for $T > 1500$ K is designed to mimic the effect of grains sublimating.
The timescale for dust to reach thermal equilibrium is extremely
short and, thus, we assume it to be in instantaneous equilibrium.  We
calculate the dust temperature by solving Equation \ref{eqn:tdust} for
$T_{\rm gr}$ using Newton's method.  From this solution, a corresponding
heating/cooling term is added to the gas following Equation
\ref{eqn:gasdust}.  The dust temperature is then used to calculate the
rate coefficient for H$_{2}$ formation on dust, which is a function of
both the gas and dust temperatures as well as the number density of
grains.  The exact form of this rate is given by
\citet{2000ApJ...534..809O}, who derive it from that of
\citet{1985ApJ...291..722T}.  This reaction can be extremely
efficient and the heating resulting from molecule formation can
significantly heat the gas if the total H$_{2}$ binding energy is
returned to the gas.  For this reason, we also include the
appropriate chemical heating term, as described in Section
\ref{sec:chemheat}.  We note that the amount of H$_{2}$ binding
energy that goes toward heating the gas is highly uncertain
\citep[see e.g.,][for a brief review of theoretical and experimental
efforts to determine the fraction of energy going into
heating]{2010ApJ...725.1111I} and so we
adopt this option as it is the most commonly employed.
We do not account for heating of the dust grains
due to H$_{2}$ formation on their surfaces, as this effect is minor compared 
to the other terms in Equation \ref{eqn:tdust}.

\subsection{Constant Heating Rates}
\label{section:constant-heating}

Additionally, the user has the option to supply arrays of constant
heating rates that will be added to the total heating/cooling rate
of each computational element due to the processes described above.
These heating rates can be either volumetric (units of
erg/s/cm$^{3}$) to mimic heat input from a radiation field or
specific (erg/s/g), corresponding to a uniform temperature change that
is independent of density.

% =============== Radiation =================

\section{Radiation Backgrounds}
\label{section:radback}

The Universe was reionized during the epoch of $z \sim 6-10$ by the
buildup of radiation from stars and active galactic nuclei (AGN).
This radiation heated the intergalactic medium (IGM) to $\sim 2 \times
10^{4}$ K \citep[e.g.][]{2000MNRAS.318..817S}, inhibiting the collapse
of halos with virial temperatures below this.  Reproducing these
effects directly in simulations requires large box sizes, extremely
high resolution, as well as radiative transfer, and is, thus,
prohibitively expensive.  A simpler approach is to make use of a
spatially uniform, redshift dependent model for the evolution of UV
background radiation, such as those introduced by
\citet{1996ApJ...461...20H}.  These models produce time/redshift
dependent spectra from which photo-heating and photo-chemical reaction
rates can be derived.  These rates can then be used in the reactions
shown in Tables \ref{tab:six} and \ref{tab:nine}.  More simply, the
spectra from UV background models can be used as inputs to
photoionization codes, like \texttt{Cloudy}, to calculate the
heating/cooling rates as a function of density, temperature, and
redshift.

As discussed in Section \ref{sec:cooling-tables}, \texttt{Grackle} makes use of
data files which store tables of all relevant chemistry and cooling
rates as a function of redshift for each UV background model.  For the
six and nine species primordial chemistry networks, we store
photo-ionization heating rates for H, He, and He$^{+}$; photo-ionization rates
for H, He, He$^{+}$, and H$_{2}$; photo-dissociation rates for H$_{2}$ and
H$_{2}^{+}$; and the photo-detachment rate for H$^{-}$.  For the tabulated
cooling method, we store the total heating and cooling rates for the
primordial and metal species as well as the mean molecular weight.
These tables are also functions of density and temperature as they are
created under the assumption of ionization equilibrium.

Currently, two UV background models are available for use with
\texttt{Grackle}.  These are the models of \citet{2009ApJ...703.1416F} and
\citet{2012ApJ...746..125H}.  Data tables for new models can be
created following the method described in Section
\ref{sec:cooling-tables}.

\subsection{Approximate Self-shielding of the UV Background}
\label{section:UVB-self-shielding}
Self-shielding against the UV photoionizing background can be
important in many applications. However, accounting for this effect
directly requires full radiative transfer, which is often
computationally infeasible. In many cosmological simulations, the UV
background is commonly taken to be optically thin everywhere, which may not always be an appropriate assumption. If desired, the user may include one of three analytic self-shielding prescriptions which operate independently on each computational element. Each method stems from the analytic fits to radiative transfer simulations from \citet{2013MNRAS.430.2427R}. They found that H self-shielding occurs at densities
\begin{eqnarray} \label{eq:nssh}
n_{\rm{H,SSh}} & \sim & 6.73\times10^{-3} \rm{cm}^{-3} \left(\frac{\bar{\sigma}_{\nu}}{2.49\times10^{-18} \rm{cm}^{2}}\right)^{-2/3} \nonumber \\
 & & \times ~ T_{4}^{0.17} \Gamma_{-12}^{2/3} \left(\frac{f_{\rm{g}}}{0.17}\right)^{-1/3},
\end{eqnarray}
where $\bar{\sigma}_{\nu}$ is the gray (spectrum-averaged) absorption cross-section, $T_{4} = T / 10^{4} \: {\rm K}$, $\Gamma_{-12}$ is the photoionization rate in units of $10^{-12} \: {\rm s^{-1}}$, and $f_g$ is the absorber baryon fraction, which we take as $f_g = 0.17$ for simplicity. Both the ionization and photoheating rates are then attenuated due to self-shielding by a factor:
\begin{equation} \label{eq:gamma_shield}
\frac{\Gamma_{\rm{shield}}}{\Gamma_{\rm{UVB}}} = 0.98 (1 + x^{1.64})^{-2.28} + 0.02 (1 + x)^{-0.84},
\end{equation}
where $x = n_{\rm H} / n_{\rm H,SSh}$.

All three available methods are various applications of equations \ref{eq:nssh} and \ref{eq:gamma_shield}. The first includes self-shielding in H only by applying these equations, leaving He and He$^+$ optically thin. The second includes self-shielding in both neutral H and neutral He using these equations, leaving He$^+$ optically thin. Finally, the third applies these equations for both neutral H and He as before, while ignoring He$^+$ photoionization/photoheating from the UV background entirely. (In other words, when accounting for self-shielding, leaving He$^+$ optically thin to the UV background may be much worse than ignoring it entirely). The latter is a common simplifying assumption in radiative transfer simulations for the H reionization epoch (but not during He reionization!) that is generally found to be a reasonable approximation \citep{2006agna.book.....O, 2010MNRAS.408.1945M, 2012MNRAS.421.2232F, 2013MNRAS.430.2427R}.  

By default self-shielding is off; these methods should be used with care, as these equations may not be applicable in all situations. This is particularly true in regimes where the ionization rate becomes dominated by collisional ionization \citep{2013MNRAS.430.2427R}, as is the case at high densities or for low UV background ionizing rates.  Finally, we note that no shielding correction is applied to the metals, which can cause large errors in the cooling due to the very different predicted electron fractions (in the non-equilibrium calculation vs.\ the \texttt{Cloudy} metal tables). Great care must therefore be taken when using this feature with metal cooling.  We note that it is, in principle, possible to address this shortcoming by generating \texttt{Cloudy} tables including shielding effects.

While equations \ref{eq:nssh} and \ref{eq:gamma_shield} are redshift independent fitting formulae, the gray-averaged cross section, $\bar{\sigma}_{\nu}$, depends on the evolving spectrum of the UV background. Included in the \texttt{Grackle} data files are the pre-computed $\bar{\sigma}_{\nu}$ for H, He, and He$^+$ at each redshift for both the \citet{2009ApJ...703.1416F} and \citet{2012ApJ...746..125H} UV background models using the frequency dependent photoionization cross sections from \citet{1996ApJ...465..487V}.\footnote{Source code containing the analytic fits given in \citet{1996ApJ...465..487V} was obtained from \texttt{http://www.pa.uky.edu/$\sim$verner/photo.html}}

\subsection{Ionization and Heating from Radiative Transfer Simulations}
\label{section:radiative-transfer}

Although \texttt{Grackle} itself does not perform radiative transfer, it can be used with a simulation code that does.  In particular, we allow for the optional inclusion of (spatially-varying) arrays of H, He, and He$^+$ photo-ionization rates, as well as a $\rm{H}_2$ photo-dissociation rate, from radiative transfer calculations. Associated heating from these processes is handled through a single heating rate array. These rates are included for each computational element and are tied to the overall heating/cooling and chemistry rates. This allows the user to couple radiative transfer solutions self-consistently with the chemical reaction network.  This could be done as a post-processing step, usually for cosmic reionization calculations \citep[e.g.][]{2014MNRAS.439..725I, 2016MNRAS.459.2342M}, or coupled with the hydrodynamics, which is becoming more commonplace in galaxy and star formation simulations \citep[e.g.][]{2014MNRAS.442.2560W, 2015MNRAS.451...34R,  2015arXiv151100011O, 2015MNRAS.454..380B, 2016arXiv160300034P, 2016arXiv160703117R}.  Our current interface only connects to primordial rates, but in the future, additional connections to radiative transfer models could include a more accurate computation of (i) the photoelectric effect, or (ii) the heating and cooling rates from metals, where the local UV/X-ray flux would be an additional interpolation variable in the lookup table \citep[e.g.][]{2013ApJ...771...50A}.

\section{Implementation} \label{methods:code}

In this section, we discuss details of the code itself, including the
available application program interfaces (APIs), code layout, and how
the existing models and networks may be extended.  For more detailed
information, users should consult online documentation available at
\texttt{https://grackle.readthedocs.org}.  In addition to the
documentation, the \texttt{Grackle} community maintains a mailing list
where users can post questions or comments and receive help from other
users and developers.  More information can be found on the front page
of the documentation.

\subsection{Simulation Code API} \label{methods:api}

The \texttt{Grackle} library provides five main functions to the user
for use in simulation codes through C or Fortran bindings: solving the
chemistry and cooling (i.e., updating the chemical species and
internal energy) and calculating the cooling time, temperature,
pressure, and ratio of specific heats.  Before these functions can be
called, the code must be initialized with various user-specified
settings.  This initialization process is also responsible for loading
data from external files and calculating the chemistry and cooling
rate tables used by the solvers.  All \texttt{Grackle} run-time
parameters are stored within a C \texttt{struct} of type
\texttt{chemistry\_data}.  The user initializes this structure by
calling the function \texttt{set\_default\_chemistry\_parameters} and
supplying a pointer to a \texttt{chemistry\_data} structure.  The
\texttt{chemistry\_data} pointer is then attached to a globally
viewable pointer called \texttt{grackle\_data}, allowing all run-time
parameters to be accessible without having to store the struct
manually.  Once all parameters are properly set, the user must call
\texttt{initialize\_chemistry\_data} to finalize the initialization
process.  An example of this procedure is shown below:

\vspace{0.5cm}
\begin{minipage}[b]{0.5\linewidth}
\begin{verbatim}

int rval;
chemistry_data my_pars;
rval =
  set_default_chemistry_parameters(&my_pars);

grackle_data.use_grackle = 1;
grackle_data.with_radiative_cooling = 1;
grackle_data.primordial_chemistry = 3;
grackle_data.metal_cooling = 1;
grackle_data.UVbackground = 1;
grackle_data.grackle_data_file = 
  "CloudyData_UVB=HM2012.h5";

rval = initialize_chemistry_data(&my_units);

\end{verbatim}
\end{minipage}

In the above example, the variable \texttt{my\_units} is a C
\texttt{struct} that holds unit conversions from internal code units
to the CGS unit system for quantities such as density, length, and
time.  These are required in order to set up the internal unit system
for the chemistry and cooling rate tables.

Once \texttt{Grackle} has been initialized, the functionality described
above can be called by the simulation code.  The available functions
are: 1) \texttt{solve\_chemistry} to integrate chemistry and cooling
equations over a specified time step, 2)
\texttt{calculate\_cooling\_time} to calculate the cooling time
($e/(de/dt)$) for each computational element, 3)
\texttt{calculate\_temperature} to calculate the temperature from the
internal energy and chemical species densities, 4)
\texttt{calculate\_pressure} to calculate the gas pressure, and 5)
\texttt{calculate\_gamma} to calculate the ratio of specific heats.
In \texttt{Grackle}, the ratio of specific heats is only altered from that
of a monatomic gas by the presence of H$_{2}$.

For efficiency, \texttt{Grackle}'s functions are designed to operate on
multiple computational elements simultaneously.  The user provides
arrays of the required fields to \texttt{Grackle} and their values are
updated by the chemistry and cooling solvers.  Because the number of
required fields depends on the specific solver being used, \texttt{Grackle}
makes use of another C \texttt{struct} as a means of passing field
arrays to the \texttt{Grackle} functions.  The \texttt{grackle\_field\_data}
\texttt{struct} contains pointers to which can be attached the arrays
of density, internal energy, and any optional fields, such as
individual species densities or the arrays of constant heating rates
(Section \ref{section:constant-heating}).  Since arrays of the
optional fields are only accessed based on run-time parameter
settings, the user has the option of only providing the fields they
wish to use.  The field arrays can be one, two, or three-dimensional,
allowing both Lagrangian particle-based codes and Eulerian mesh-based
codes to provide fields in their native layout.  The
\texttt{grackle\_field\_data} \texttt{struct} contains entries to
specify the field dimensionality as well as to flag certain array
elements as boundary cells which are to be ignored.  An example of
calling the main chemistry solver function on a 10$^{3}$ grid with 3
boundary zones on each side is shown below:

\vspace{0.5cm}
\begin{minipage}[b]{0.5\linewidth}
\begin{verbatim}

grackle_field_data my_fields;

my_fields.grid_rank = 3;
my_fields.grid_dimension = new int[3];
my_fields.grid_start = new int[3];
my_fields.grid_end = new int[3];
for (int i = 0;i < 3;i++) {
  my_fields.grid_dimension[i] = 10;
  my_fields.grid_start[i] = 3;
  my_fields.grid_end[i] = 12;
}

my_fields.density = density_array;
my_fields.internal_energy = energy_array;
my_fields.HI_density = HI_array;
...

// 1 Myr in internal units
double dt = 3.15e7 * 1e6 /
  my_units.time_units;

rval =
  solve_chemistry(&my_units, &my_fields, dt);

\end{verbatim}
\end{minipage}

An added benefit of this approach is that adding new
features which use additional fields will not require a change in the
function signature.  This will, in theory, allow \texttt{Grackle} to maintain
backward compatibility indefinitely.  We note that versions of \texttt{Grackle}
prior to 3.0 did not make use of the \texttt{grackle\_field\_data}
\texttt{struct} and instead required all fields to be provided as
individual arguments.  We acknowledge that the release of
\texttt{Grackle} 3.0 constitutes a significant change to the API, but
one that will ultimately provide more stability moving forward.

\subsection{Pygrackle - Grackle's Python API}

As described above, \texttt{Grackle}'s native API is provided through C and Fortran
bindings.  This is particularly useful for simulation codes, as they are
typically written in either C or Fortran compatible languages, but it provides
a barrier to entry for experimental work and testing of \texttt{Grackle} functionality.
Python is a high-level, interpreted language, increasingly used in scientific
computing both as ``glue" code as well as a mechanism for authoring production
scientific codes.  For instance, in 2016, one of the Gordon Bell Prize nominees
utilized the Python code PyFR to demonstrate extreme scaling of finite element
calculations \citep{pyfr}.

To facilitate access to \texttt{Grackle} functionality, we provide Python bindings to
it, designed to make interacting with chemical rates and rate equations
available to researchers at all stages.  This enables rapid iteration over
different rate coefficients, different initial state vectors, and over
arbitrary time periods.  The bindings are written in Cython
\citep{behnel2010cython} with minimal overhead from Python operations.  Below,
we describe two particular aspects of \texttt{Pygrackle}.

\subsubsection{Fluid Container} \label{sec:pyfluid}

\texttt{Pygrackle} provides an object called a ``fluid container''.  When data is passed
to \texttt{Grackle} during the course of a simulation, it may be sent as a single zone,
as multiple zones that are organized in a 3D array, or as a 1D ``pencil beam'' of
data.  The fluid container object is designed to mimic this, enabling
individuals to ``create" a collection of fluids either by reading them from
disk or constructing them in-memory from NumPy arrays \citep{numpy}.  This
enables calculations of cooling time, chemical evolution, etc., from
analytically-defined gas parcels and distributions.  While this will not take
into account hydrodynamics (unless a package spiritually similar to \texttt{Grackle},
but for hydrodynamics, is released) it can provide useful and scientifically
relevant information.

The fluid container provides several high-level functions, such as computing
the cooling time, the pressure, and so forth, most of which are usually used
only internally in \texttt{Grackle}.  Additionally, this object provides compatibility
with \texttt{yt} \citep{2011ApJS..192....9T}, enabling individuals to load data
with \texttt{yt} and then evolve it through \texttt{Grackle}.  One possible use case for
this is to load in a dataset and use \texttt{Pygrackle} to compute different cooling
times for a collection of gas identified in \texttt{yt} based on different
metallicity assumptions, different chemical rate coefficients, and different
radiation backgrounds.

\subsubsection{Evolution Models} \label{sec:pyevolve}

In addition to providing access to \texttt{Grackle}'s primary
functionality, \texttt{Pygrackle} also features a set of convenience functions
to evolve a fluid container forward in time following simple models.
These functions return a dictionary of arrays of all fluid quantities
(i.e., species densities, internal energy, temperature, etc.) for time
values of the evolution.  These functions can be used in semi-analytic
models that require knowledge of the thermal evolution of gas under
different conditions.  Examples of use are discussed in
Section \ref{sec:testing}.

The simplest of these evolves a fluid container assuming a constant
density model.  The \texttt{evolve\_constant\_density} function takes
an initialized fluid container and repeatedly calls \texttt{Grackle}'s
\texttt{solve\_chemistry} function until a specified stopping time or
temperature has been reached.  For each iteration, the timestep is
taken to be a fraction of the local cooling time, specifiable by the
user and defaulting to 0.01.

The second of these functions models the evolution of a parcel of gas
collapsing due to self-gravity.  The \texttt{evolve\_freefall} function
closely follows the one-zone free-fall collapse model introduced by
\citet{2000ApJ...534..809O} and modified by
\citet{2005ApJ...626..627O} to include the effects of thermal pressure
support.  The gas density evolves following the modified collapse
model of \citet{2005ApJ...626..627O}, given by
\begin{equation}
\frac{d \rho}{dt} = \frac{\rho}{t_{\rm col}},
\end{equation}
where $t_{\rm col}$ is the collapse time-scale expressed as
\begin{equation}
t_{\rm col} = \frac{t_{\rm ff}}{\sqrt{1-f}},
\end{equation}
and the free-fall time is given by
\begin{equation}
t_{\rm ff} = \sqrt{\frac{3 \pi}{32 G \rho}}.
\end{equation}
Thermal pressure forces, which act to slow the collapse of the cloud,
are modeled by the factor $f$, which is expressed as
\begin{equation}
f = \left\{
\begin{array}{lr}
0, & \gamma < 0.83,\\
0.6 + 2.5(\gamma - 1) - 6.0(\gamma - 1)^2, & 0.83 < \gamma < 1,\\
1.0 + 0.2(\gamma - 4/3) - 2.9(\gamma - 4/3)^2, & \gamma > 1,
\end{array}
 \right.
\end{equation}
where the effective adiabatic index, $\gamma$, is
\begin{equation}
\gamma \equiv \frac{\partial \ln p}{\partial \ln \rho}.
\end{equation}
As the density increases, the internal energy is altered by a
combination of adiabatic compression and radiative cooling, computed
by \texttt{Grackle}, and is given by
\begin{equation}
\frac{de}{dt}= -p \frac{d}{dt} \frac{1}{\rho} - {\Lambda},
\label{eq:energy}
\end{equation}
where the pressure is given by
\begin{equation}
p = \frac{\rho k T}{\mu m_{\rm H}},
\end{equation}
the specific internal energy is given by Equation \ref{eqn:e-T},
and $\Lambda$ is the
radiative cooling rate.  For each iteration of this function, the
timestep is taken to be a fraction of the local free-fall time,
specifiable by the user and defaulting to 0.01.  An example use of
this function is shown in Section \ref{sec:free-fall-test}.

%\subsection{Code Structure} \label{Code_Structure}

\subsection{Adding New Models/Rates}

Adding new rates to \texttt{Grackle} involves modifying the values stored in
the rate coefficients and additionally defining a new network for both
the chemistry and if desired also the cooling. As it stands, some
direct modification of the code structure is required (although see \S
\ref{Future_Directions} for future improvements in this area), and we
give more details of that in this section.

The structure of the current code base is set up to realize either an
equilibrium chemistry model using cooling tables derived from \texttt{Cloudy},
or a non-equilibrium cooling model based on a six, nine or twelve
species chemistry model.  To implement a different equilibrium cooling
model, the process is relatively straightforward: \texttt{Grackle} reads
the \texttt{Cloudy} cooling and interpolates the data along one, two or three
dimensions as appropriate. By substituting in a different cooling
table, with the correct format, the behavior of the cooling for a
given species can be easily changed.

To modify the non-equilibrium cooling behavior, more work is
required. \texttt{Grackle} assumes that non-equilibrium chemistry models are
hierarchical, with the nine species model composed of the six species
model plus three additional species and likewise for the twelve species model.
To expand the network to include, say, a fourth model containing the twelve species 
model as a subset then the procedure is straightforward. However, if a different
chemical network is envisaged containing some species already in the
twelve species model and some not, then this will require modifying
the existing hierarchical structure.  The next sections provide some
details of this process.

\subsubsection{Updating the rate coefficients}

The rates are allocated and declared in the
\texttt{initialize\_chemistry} function. To update the rate
coefficients (e.g., based on more recent experimental data), simply
make use of the already existing rate coefficient arrays (\texttt{k}
array) as declared in \texttt{initialize\_chemistry}. The rates are
populated in \texttt{calc\_rates\_g}. Within this function the rates
are fully described. For example, \texttt{k1} is the rate coefficient
for the collisional ionization of neutral hydrogen by electrons
(\texttt{k1}: H + e$^{-}$ $\rightarrow$ H$^{+}$ + 2e$^{-}$). If a
significant number of new rate coefficients are needed then the most
expedient approach would be to insert a preprocessor directive into
\texttt{calc\_rates\_g} in which the appropriate function call can be
inserted.  For any additional rate coefficients that are required, the
corresponding \texttt{k} array needs to be declared and allocated in
\texttt{initialize\_chemistry}.  Furthermore, the interpolation of the
new rates will need to be added in the function
\texttt{lookup\_cool\_rates1d}.

\subsubsection{Updating the chemistry network}

To update the chemical network to either include or exclude reactions,
a new rate network will be required. As a template, the function
\texttt{step\_rate} can be used.  If the new network is simply an
addition to the existing network (e.g., a 15 species model) then the
easiest option is to simply augment this network with the three extra
species using the appropriate interactions. The species will then be
evolved until they converge. More complex additions would require
creating a new network update routine using \texttt{step\_rate}
as a template.

\subsubsection{Updating the cooling model}

Apart from the chemical network, the cooling model may also be
modified. As discussed previously, if the intention is to implement a
new cooling table then the changes are straightforward.  For changes
to the equilibrium network (say, to modify the cooling rate due to, for
example, emission line cooling from a given species), this is handled
in \texttt{cool1d\_multi\_g}. As discussed in \S \ref{Cooling}, line
emission cooling is determined using collisional excitation,
collisional ionization and recombination rates. If the intention is to
update/modify existing rates then the cooling rates are also set in
\texttt{calc\_rates\_g} and can be modified there. If new cooling
rates are required from another species whose cooling properties are
important for the network then the rates can be added there also. Any
new arrays required in this case will also need to be declared and
initialized in \texttt{initialize\_chemistry} in a way similar to the
rate coefficients. The rates can then be interpolated to the required
temperature values in \texttt{calc\_rates\_g}, following the examples
there. Finally, once the new cooling rate has been determined its
values needs to be added to the \texttt{edot} array, which tracks
the total cooling/heating rate.

\section{Profiling and Testing}
\label{sec:profiling-and-testing}

\subsection{Testing Framework}
\label{sec:testing}

Testing a library like \texttt{Grackle}, with its mix of \texttt{Fortran}, and
\texttt{C++} code, can be difficult. In the interest of ultimately improving
test coverage and making it easier to prototype tests, \texttt{Grackle} employs a
Python-based testing framework that makes heavy use of the \texttt{Pygrackle}
Python wrapper for \texttt{Grackle}. The tests are orchestrated using the
\texttt{py.test}\footnote{http://pytest.org/} package.

Currently there are two different types of tests in the \texttt{Grackle} test suite: unit
tests and answer tests. A unit test in \texttt{Grackle} compares the results of a
calculation using \texttt{Grackle} to some set of ``correct'' answers that are known
\textit{a priori}. The unit tests currently implemented in \texttt{Grackle} check that the
unit system is behaving correctly (see Section~\ref{sec:test-units}) and that the
ionization equilibrium for a primordial gas agrees with the analytic prediction
using the rates implemented in \texttt{Grackle} (see Section~\ref{sec:test-cie}).

The answer tests consist of a set of example calculations where each calculation
writes out a summary plot as well as an \texttt{HDF5} dataset that is loadable
by the \texttt{yt} package. Known ``correct'' answers for the summary plot and
\texttt{yt}-loadable dataset are saved in the repository so that code changes in
\texttt{Grackle} can be tested to ensure that answers produced by the library do not
change. This process does not prevent \textit{incorrect} answers from being
generated initially, but it does notify the developer if answers change
as a result of a code modification. Incorrect answers are prevented by manually inspecting test
answers when the test is first added to the codebase. If subsequently a bug is
discovered (or an enhancement to the code is made) and the test output changes, then the test answer must also be
manually updated. Currently, \texttt{Grackle} contains stored answers for: the
instantaneous cooling rate (see Section~\ref{sec:cooling-rate-test})
at a constant density, the temperature evolution of a uniform-density cloud (see
Section~\ref{sec:uniform-cooling-test}), and the density and
temperature evolution of a gas cloud undergoing free-fall collapse
(see Section~\ref{sec:free-fall-test}). The answer tests are run
several times using different input physics to ensure \texttt{Grackle}'s
solvers are well-exercised by the tests.  The answer tests are
presented as sample scripts that can be run manually by the user,
producing a figure.  For each answer test, we show the corresponding
figure exactly as produced by each script.

In addition to the unit and answer tests, which monitor functionality
of the solvers, we also employ tests to ensure that all Python source
code conforms to PEP 8 style
standards\footnote{https://www.python.org/dev/peps/pep-0008/} and that
all instructional sample codes compile and run without producing
errors.

\subsubsection{Unit Test: Unit Systems}
\label{sec:test-units}

For a given set of physical conditions, (i.e., densities and internal
energies), the results of \texttt{Grackle}-related calculations should be
independent of the choice of reference frame (comoving or proper) and
internal unit system.  However, because chemistry and cooling
calculations involve numerical values that span many orders of
magnitude, round-off error will eventually lead to significant
differences when the internal unit systems are varied beyond a certain
degree.  The unit systems unit tests set up two fluid container
objects with the same physical conditions but different internal unit
systems.  In each instance, the chemical species fractions are evolved
until ionization equilibrium has been reached (see \S
\ref{sec:pyfluid}), after which time the cooling time is calculated.
The tests assert that the cooling time values agree to within 4
decimal places between the two unit systems.

Three variants of this unit test exist.  In the first two, a
comoving-frame unit system appropriate for a cosmological simulation
is compared with a proper-frame unit system drawn from a random number
generator that allows the density, time, and length units to vary by 4
orders of magnitude.  A cosmologically appropriate unit system is
roughly defined as one with density units equal to the average
comoving matter density of the Universe, $\bar{\rho}_{m}$; time units
proportional to 1/$\sqrt{G\ \bar{\rho}_{m}}$; and length units on the
scale of Mpc.  One of the two of these type is performed with the
non-equilibrium chemistry solver and the other with the fully
tabulated solver.  In both cases, we compare a randomly generated
proper-frame unit system with a comoving-frame unit system at $z = 0$
and $z > 0$, where the proper and comoving frames differ.  In these
tests, a UV background model is also used as the radiative heating
rate is proportional to $\rho$, whereas collisional heating/cooling
rates are proportional to $\rho^{2}$ (or $\rho^{3}$ for three-body
reactions).  Including heating/cooling terms with different density
scaling is useful for exposing errors in adjusting between comoving
and proper reference frames.

The final variant of the unit system test compares two randomly
generated proper-frame unit systems whose density units differ by as
much as possible while maintaining equivalency to 4 decimal places.
In practice, we find that the density units can differ by roughly 31
orders of magnitude before the threshold level of accuracy is lost.
With this in mind, we allow the two randomly generated unit systems to
span 2 orders of magnitude with the center of each random distribution
chosen such that the unit system will differ by 27-31 orders of
magnitude.  By comparing unit systems that differ by the maximally
allowed amount, we are able to measure the degree to which new terms
added to the network suffer from round-off error.

\subsubsection{Unit Test: Collisional Ionization Equilibrium}
\label{sec:test-cie}

The equilibrium ionization state of a gas is determined solely by its
temperature when only collisional ionization is considered, (i.e.,
when photo-ionization is neglected).  Thus, a density-independent
equilibrium solution can be calculated for any ion by equating the
creation terms (recombination from a higher ionization state and 
ionization from a lower state) with the destruction terms
(recombination into a lower state and ionization into a higher
state).  For example, this yields a CIE solution for neutral H given
by
\begin{equation} \label{eqn:hcie}
f_{\rm H,CIE} = \frac{\alpha_{\rm H^{+}}(T)}{\alpha_{\rm H^{+}}(T) +
  \Gamma_{\rm H}(T)},
\end{equation}
where $\alpha_{\rm H^{+}}$ is the recombination rate of H$^{+}$ and
$\Gamma_{\rm H}$ is the collisional ionization rate of H.

In order to test that \texttt{Grackle} arrives at the correct CIE solution for
the atomic primordial network, we initialize a fluid container at a
constant density with temperature varying smoothly in log-space from
10$^{4}$ K to 10$^{9}$ K.  The gas is initialized in a fully neutral
state and the chemistry solver is called repeatedly with cooling
processes deactivated (to keep each cell at its original temperature)
until convergence has been reached in all cells.  These values are
then compared to the analytical solutions (as in Equation
\ref{eqn:hcie}) for the ionization states of all H and He species and
the total electron fraction.

\subsubsection{Answer test: cooling rate test}
\label{sec:cooling-rate-test}

\begin{figure}
  \centering
  \includegraphics[width=0.45\textwidth]{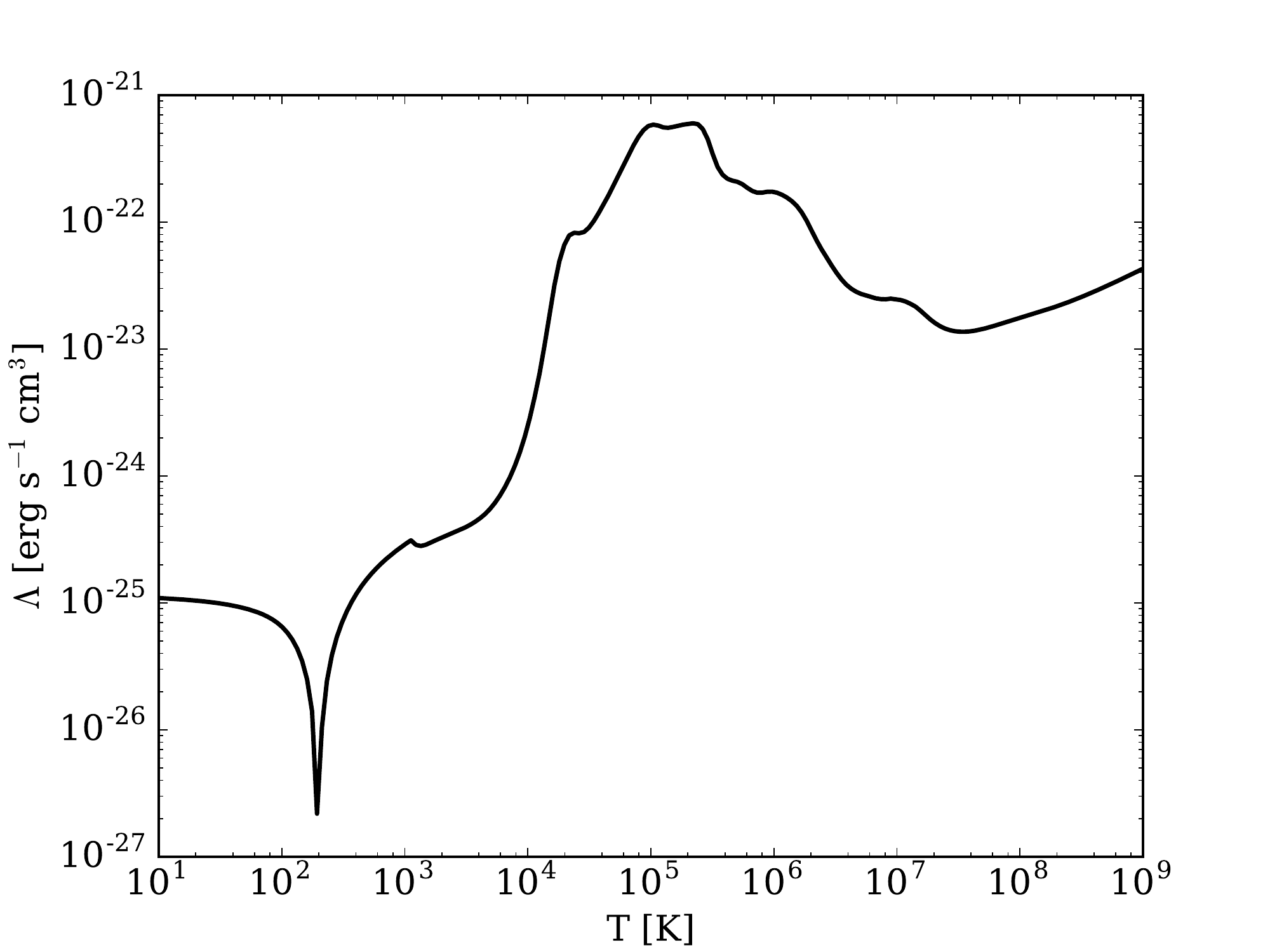}
  \caption{ Figure output by the default configuration of the cooling
    rate answer test, described in Section
    \ref{sec:cooling-rate-test}, showing the equilibrium cooling rate
    as a function of temperature for a Solar metallicity gas at a
    density of 1 amu/cm$^{3}$ with an incident radiation field
    described by the \citet{2012ApJ...746..125H} model at z = 0.  The
    absolute value of the cooling rate is shown (in order to use a
    log scale) because below $\sim$ 200 K, the radiation field
    induces a net heating rate.  } \label{fig:cooling-rate-test}
\end{figure}

Similar to the CIE unit test (Section \ref{sec:test-cie}), the cooling
rate answer test initializes a fluid container with a constant density
and smoothly varying temperature from 10 K to 10$^{9}$ K, then
iterates the chemistry solver until all species have reached
equilibrium.  Unlike the CIE unit test, metal cooling and a
\citet{2012ApJ...746..125H} UV background at $z = 0$ are also included.
After reaching equilibrium, the total cooling rate is calculated and
compared with stored answers, as described in Section
\ref{sec:testing} above.  This test is performed for all versions of
the primordial solver as well as the fully tabulated cooling solver.
The figure produced by the default configuration of this test
(H, D, He non-equilibrium solver) is shown in Figure
\ref{fig:cooling-rate-test}.

\subsubsection{Answer test: constant density cooling test}
\label{sec:uniform-cooling-test}

\begin{figure}
  \centering
  \includegraphics[width=0.45\textwidth]{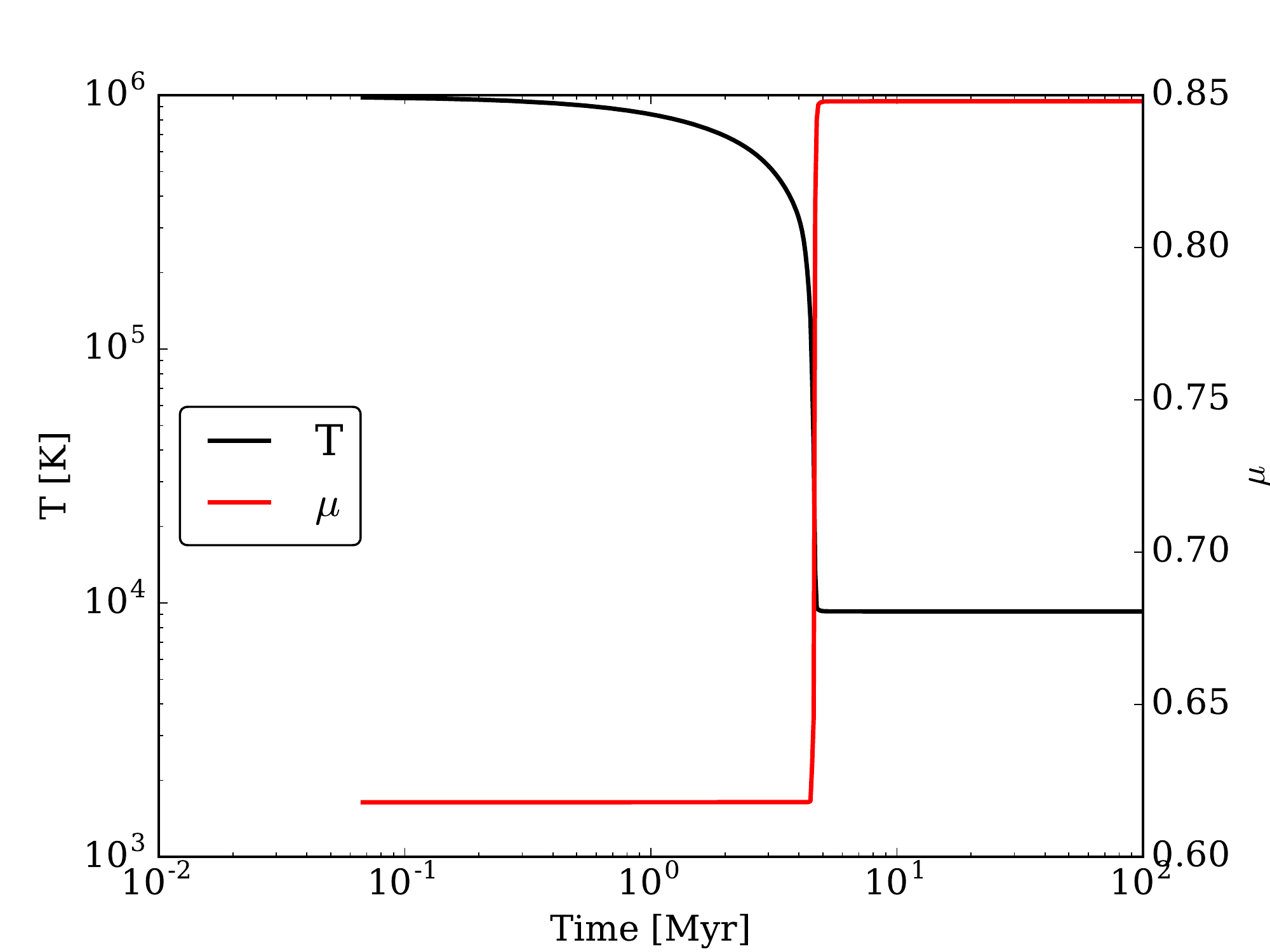}
  \caption{
    Figure output by the uniform cooling answer test, described in
    Section \ref{sec:uniform-cooling-test}, showing the temperature
    (black) and mean molecular weight (red) as a function of time for
    a parcel of gas cooling at constant density.
  } \label{fig:uniform-cooling-test}
\end{figure}

The uniform cooling answer test simulates the thermal evolution of a
parcel of gas cooling at constant density.  This test ensures that the
solver properly evolves the internal energy over a period of time.
The test initializes a single-cell fluid container with a density of
0.1 amu/cm$^{3}$ and a temperature of 10$^{6}$ K.  The cell is evolved
for 100 Myr using the \texttt{evolve\_constant\_density} convenience
function with timesteps of 1\% of the cooling time.  The
resulting evolution is shown in Figure
\ref{fig:uniform-cooling-test}.

\subsubsection{Answer test: free-fall collapse test}
\label{sec:free-fall-test}

\begin{figure}
  \centering
  \includegraphics[width=0.45\textwidth]{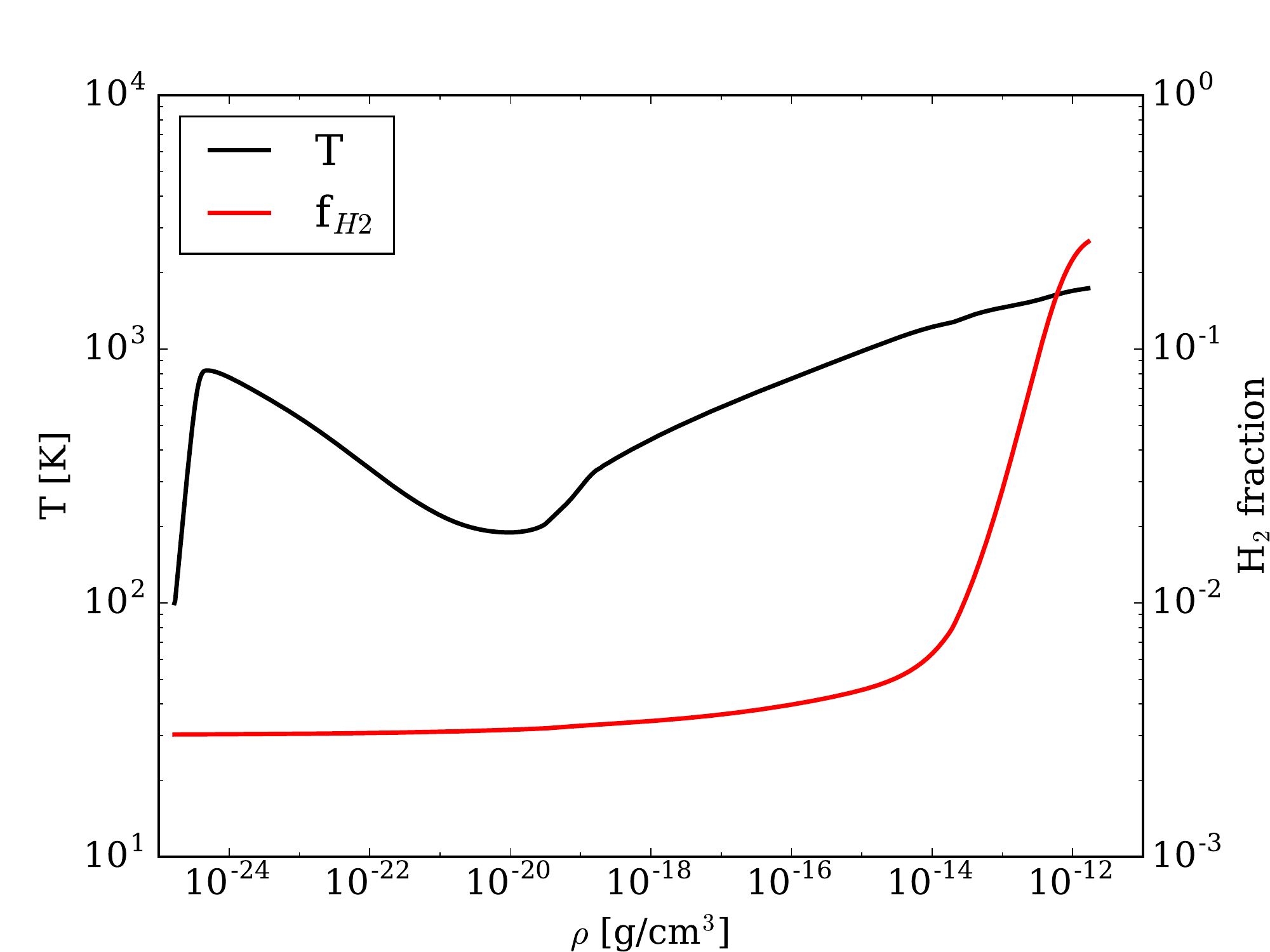}
  \caption{
    Figure output by the default configuration of the free-fall answer
    test, described in Section \ref{sec:free-fall-test}, showing the
    temperature (black) and H$_{2}$ fraction (red) as a function of
    density for a free-fall collapse model of a metal-free gas.
  } \label{fig:free-fall-test}
\end{figure}

The free-fall collapse answer test simulates the thermal evolution of
a cloud of gas collapsing due to self-gravity.  This test is useful
for ensuring that the non-equilibrium chemistry solver is functioning
properly over a large range in density.  The test creates a
single-cell fluid container with an initial density of 0.1
amu/cm$^{3}$ and an initial temperature of
50,000 K.  Before beginning the free-fall collapse phase, the cloud is
allowed to cool at a constant density to a temperature of 100 K using
the \texttt{evolve\_constant\_density} function described in Section
\ref{sec:pyevolve}.  This allows the gas to settle into an ionization
state that is roughly appropriate for the temperature.  From there, we
evolve the density of the cloud using the \texttt{evolve\_freefall}
function, also discussed in Section \ref{sec:pyevolve}.  This test is
performed using the full non-equilibrium chemistry solver, once at
zero metallicity (Figure \ref{fig:free-fall-test}) and once with a
metallicity of 10$^{-3}$ Z$_{\odot}$.

\subsection{Performance}

\subsubsection{Optimization Strategy}

Our optimization strategy in the \texttt{Grackle} has two components related to serial and parallel execution.  We begin with single processor optimization.  

The ordinary differential equations that describe chemical and thermal evolution do not use spatial information and so each discretization point (particle or cell) can be evolved independently of the others.  Because of this, the \texttt{Grackle} can be used with a wide variety of codes and applications, and optimization is relatively straightforward.  Our primary technique for single processor optimization is to make good use of cache and (single processor) vectorization.  The API is built around the idea of ``fields" of points (fluid containers) rather than a single point for this reason.  The field can be a single-dimensional contiguous list as might be used for particle-based codes, or a three dimensional grid with inactive (``ghost'') points as would be appropriate for grid-based codes.  By taking an entire field, and operating on the field in an order corresponding to the way it is laid out in memory, the code tries to minimize cache misses.  In particular, multi-dimensional arrays are assumed to be Fortran-ordered (column-major) and operations are performed in loops over the most rapidly varying index.  Loops are generally also written in a way which facilitates unrolling so that compilers can easily make use of vector operations and prefetching.  Much of the computationally intensive part of the code is written in Fortran (in part for historical reasons but in part to take advantage of well-tested Fortran compiler loop optimizations).

The second optimization involves taking advantage of parallel computation.  \texttt{Grackle} itself requires no communication (and is completely thread-safe after the initialisation step) and so can easily work as part of a code that uses MPI or some other message passing library to achieve distributed parallelization.  In addition, \texttt{Grackle} supports OpenMP parallelization and thus can easily work with applications adopting a hybrid MPI/OpenMP model.  The OpenMP is implemented by parallelizing over outer j,k loops and giving a thread an i ``slice" to operate on.  This is a natural model for structured grid-based codes, but some work may be required to get good performance this way with unstructured or particle based codes (e.g., by artificially splitting a 1D list of particle quantities into a 2D grid).

\subsubsection{Serial Performance}

\begin{figure}
  \centering
  \includegraphics[width=\columnwidth]{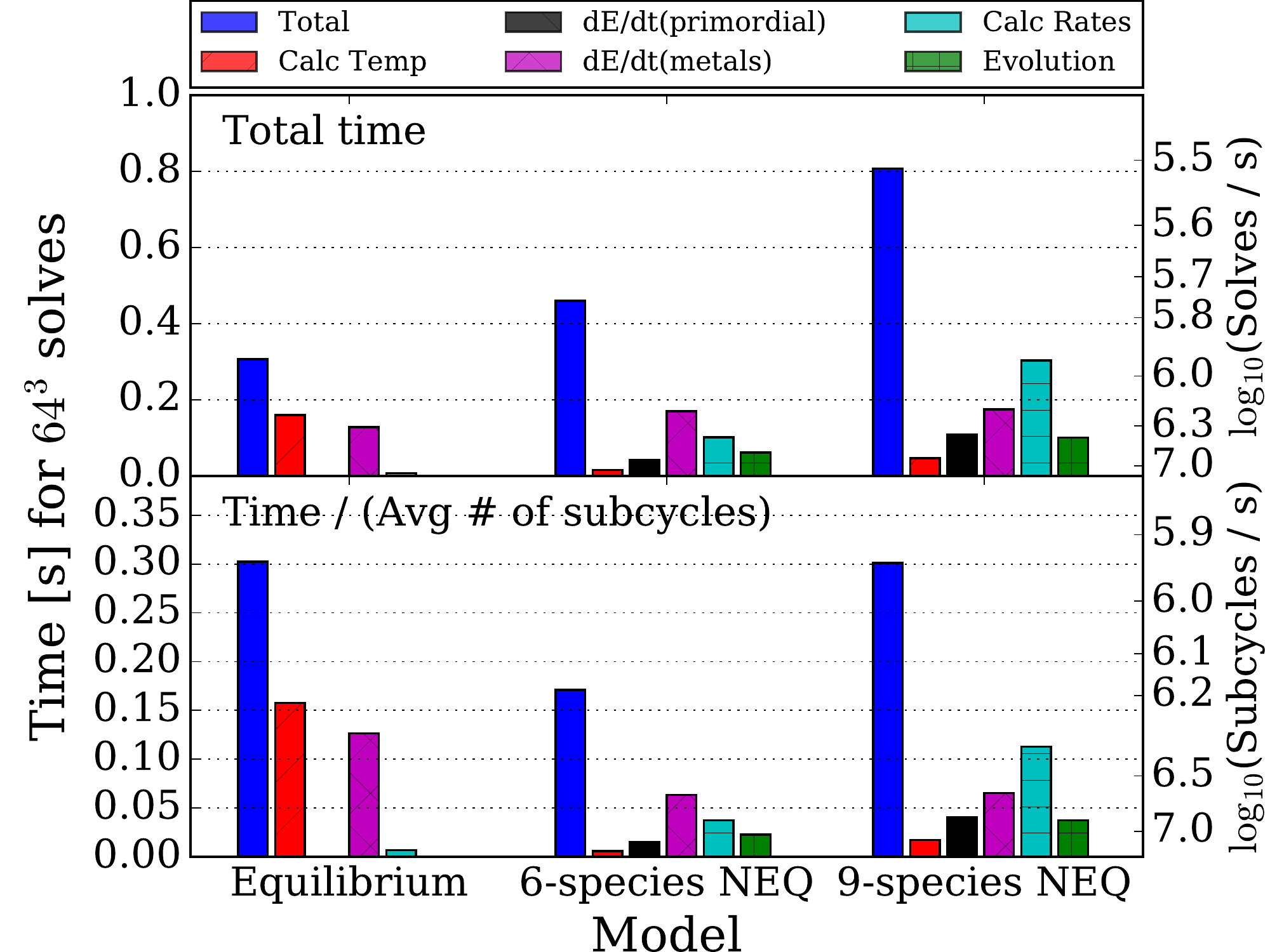}
  \caption{Top panel: Total time to compute the cooling rate test
    (\S\ref{sec:cooling-rate-test}) in $64^3$ fluid containers, using
    the tabulated equilibrium cooling model (left), six species
    non-equilibrium model (middle), and nine species non-equilibrium
    model (right).  The different bars show the time needed for the
    complete solve (blue), the temperature calculation (red), the
    $\dot{e}$ computation due to primordial chemistry (black) and
    metals (magneta), the interpolation of rate coefficients (cyan),
    and the update of the species fractions with backward
    differencing.  Bottom panel: Total time but normalized by the
    average number of subcycles per cell, which demonstrates the
    performance of a single iteration in each solver
    mode.} \label{fig:performance}
\end{figure}

We utilize the cooling rate test (\S\ref{sec:cooling-rate-test}) to
assess the performance of \texttt{Grackle}.  We set up the test with $64^3$
fluid containers with hydrogen number density $n_{\rm H}$, temperature
$T$, and metallicity $Z$ varying in each dimension.  These quantities
are equally log-spaced in the range $\log (n_{\rm H} /
\textrm{cm}^{-3}) = [-1, 3]$, $\log (T/\textrm{K}) = [1, 8]$, and
$\log (Z/Z_\odot) = [-4, 0]$.  All of the fluid containers are
initially neutral.  We run each test with the tabulated solver in
ionization equilibrium and the non-equilibrium solver with the
six species and nine species networks on a single core of a desktop
computer with dual Intel Xeon ``Westmere'' E5645 CPUs at 2.4 GHz, each
of which has six cores.  The tests are evolved for 500 yr. In
most cases, this is shorter than the cooling time, but it provides an
ample test for the performance of \texttt{Grackle}.  Because the fluid
containers are not initialized in ionization equilibrium, the first
timestep in the non-equilibrium solvers requires hundreds of subcycles
for the system to converge.  Due to the fact that the non-equilibrium solver
performance is directly tied to the number of subcycles, we do not
include the first three cycles of the tests (which are not representative of
typical use conditions).

% Something about this is an ideal solve, but performance can diminish
% if more difficult (high densities, shocks, or ionization) conditions
% exist.

The top panel of Figure \ref{fig:performance} shows the total time
(blue bars) required for this $64^3$ performance test in each solver
mode.  This is further divided into the time spent in each major
component of \texttt{Grackle}: the calculation of the temperature (red),
$\dot{e}$ from primordial (black) and metal (magenta) species, rate
coefficient interpolation (cyan), and the update of the species
fractions (green).  From a total performance standpoint, the
non-equilibrium solver in the six species and nine species primordial
models requires 50\% and 164\% more time than the equilibrium solver,
respectively.  In this test, \texttt{Grackle} can update $8.6 \times 10^5$,
$5.7 \times 10^5$, and $3.2 \times 10^5$ fluid containers per second
for the equilibrium, six species and nine species solvers, respectively.
The computational expense in the equilibrium solver is almost evenly
split between the equilibrium temperature calculation and metal
cooling rates.  The metal cooling rate and rate coefficient
interpolation consume the most compute cycles in the six species and
nine species solvers, respectively.  The temperature calculation in the
non-equilibrium solver takes relatively little computation because it
is simply calculated from the pressure and total number density with
no iterative processes.

The cooling rate test represents a fluid in many different
chemo-thermal states, which converge to some equilibrium.  However in
production simulations, there are many ``difficult'' situations, such
as high densities, strong shocks, and strong radiation fields, in
which the equations become stiff and require many subcycles to
complete an entire solve.  Therefore to better gauge the performance
of a single iteration, we show the average time elapsed {\it per
  subcycle} for the same test in the bottom panel of Figure
\ref{fig:performance}.  The equilibrium and non-equilibrium solvers
take an average of 1.01 and 2.67 subcycles per solve, respectively.
\texttt{Grackle} can perform $8.7 \times 10^5$, $1.5 \times 10^6$, and $8.7
\times 10^5$ subcycles per second for the equilibrium, six species and
nine species solvers, respectively.  Here we see that the equilibrium
solver actually requires 75\% more time per subcycle than the six species
non-equilibrium solver and is equivalent to the performance (per
subcycle) of the nine species non-equilibrium solver.  In practice, if
any cells require many subcycles to converge to a solution, the call
to \texttt{Grackle} will require more time per cell than in this ideal test,
because the total performance is entirely dependent on the total
number of subcycles performed in one solve, not the number of cells.

\subsubsection{OpenMP Performance}

\begin{figure}
  \centering
  \includegraphics[width=0.45\textwidth]{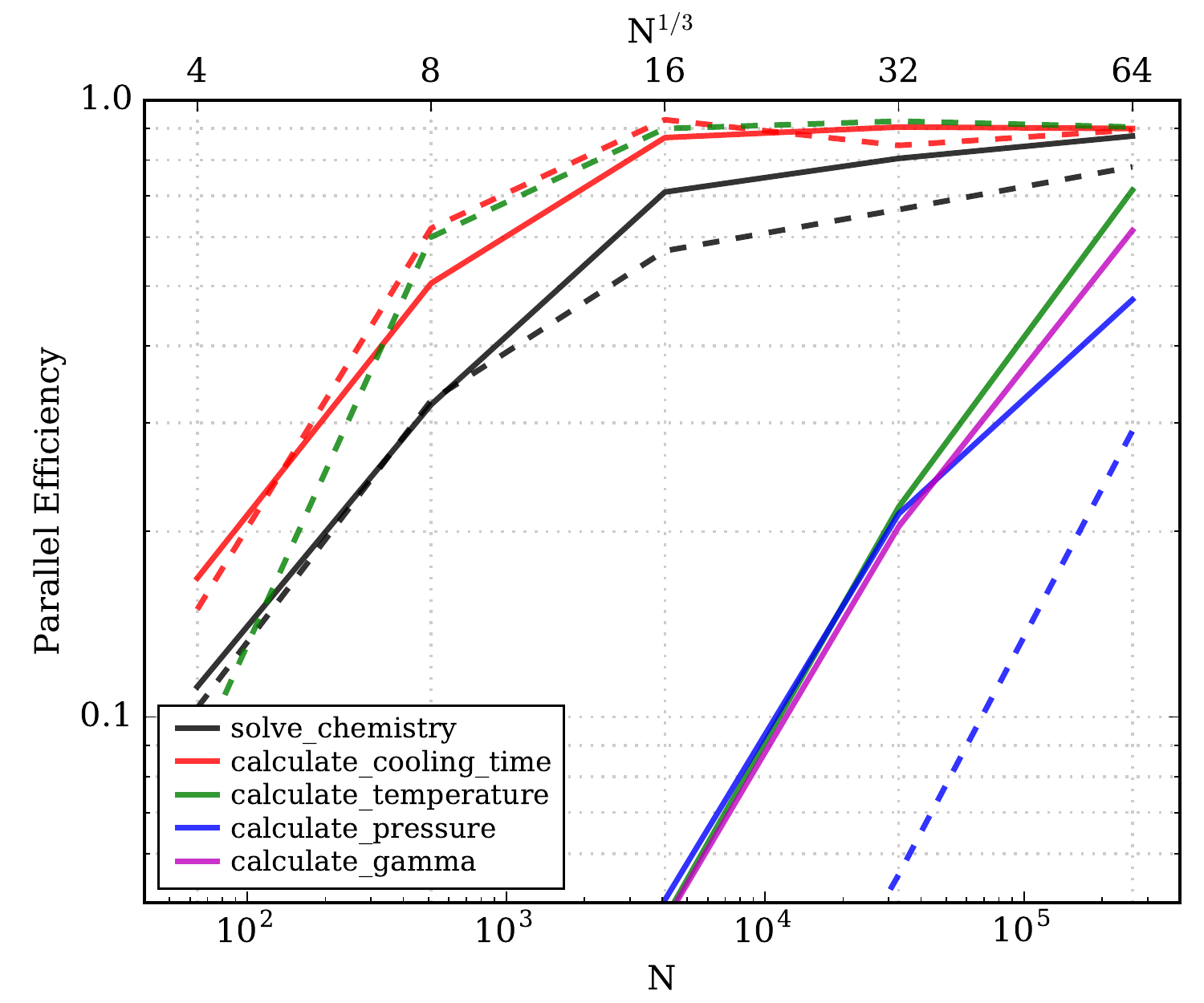}
  \caption{
    OpenMP parallel efficiency using 20 threads as a function of the size of
    the input array. The solid lines show the use of the
    non-equilibrium solver with \texttt{primordial\_chemistry} = 3 and
    the dashed lines show the analogous functions using the tabulated
    solver.  For all time-consuming
    routines, the parallel efficiency reaches $\sim$60\% to 90\% for
    $16^3$ cells and $\sim$80\% to 90\% for $64^3$ cells.
  } \label{fig:omp-perf}
\end{figure}

In addition to the single-processor performance just described, we
characterize the threaded performance of the \texttt{Grackle}.
Figure \ref{fig:omp-perf} shows an OpenMP performance benchmark for both the
non-equilibrium and tabulated functionality, where parallel efficiency is
defined as the ratio of multi-thread to single-thread performance. We
conduct this benchmark on the Campus Cluster of the University of Illinois
at Urbana-Champaign using 20 threads on two Intel Xeon E5-2670 v2 CPUs
at 2.50 GHz, each of which has ten cores. We compile \texttt{Grackle}
using version 15.0.0 of the Intel compilers with ``-O3''
optimization.  For all time-consuming routines
(i.e., calculating cooling, cooling time, and temperature with the tabulated
solver, and calculating chemistry, cooling, and cooling time with the
non-equilibrium chemistry solver), the parallel efficiency reaches
$\sim 60\%\,\text{--}\,90\%$ for $16^3$ cells and
$\sim 80\%\,\text{--}\,90\%$ for $64^3$ cells. For other computationally cheap
routines, such as calculating pressure, the parallel efficiency is relatively
low. This is not an issue since they take negligible time compared to other
computationally expensive routines.

%%% Local Variables:
%%% mode: latex
%%% TeX-master: "ms"
%%% End:

\section{Discussion} \label{sec:summary}

\subsection{Applicability and Limitations}

\begin{figure}
  \centering
  \includegraphics[width=0.48\textwidth]{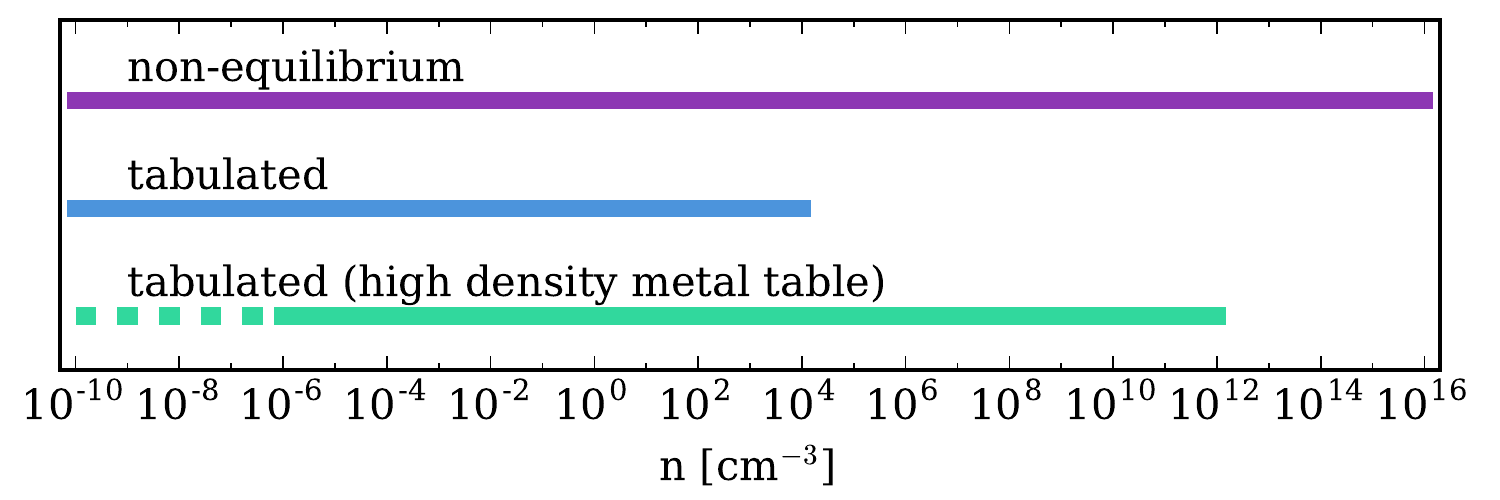}
  \caption{
    The appropriate density range for different versions of the
    \texttt{Grackle} solver.  The high density metal cooling table (bottom)
    explicitly spans the density range, 10$^{-6}$ cm$^{-3}$ $< n <$
    10$^{12}$ cm$^{-3}$, but extrapolation down to $n = 10^{-10}$ cm$^{-3}$ 
    is still valid, as indicated by the dashed line.  For each
    of these, the valid temperature range is roughly 1 K to 10$^{9}$
    K.
  } \label{fig:valid-range}
\end{figure}

It is important to remember that the range of physical conditions over
which \texttt{Grackle} can be considered valid is not unlimited.  In Figure
\ref{fig:valid-range}, we show the rough density range over which
different components of the solver are valid.  The high density limit
of the non-equilibrium solver is set roughly by the reactions present
in the network and the range over which their rate coefficients are
trusted.  The limits on the cooling tables correspond to the density
range over which they were calculated.  Users are especially
cautioned against exceeding the upper density limit of the tabulated
cooling solver.  The critical density above which energy levels are
populated according to LTE exceeds the upper limit of the tables for
many metal coolants \citep{2008MNRAS.385.1443S}.  Thus, these tables
do not capture the NLTE to LTE transition where the cooling rates
change from scaling as $\rho^{2}$ to $\rho$.  Hence, extrapolation
beyond the upper limit may result in vast over-prediction of the
cooling rate.  If the use-case requires exceeding this limit, then the
high density metal table should be used in conjunction with the
non-equilibrium solver.  In all cases, the valid temperature range is
roughly 1 K to 10$^{9}$ K.  The tables computed with \texttt{Cloudy}
are defined over this temperature range.  For the primordial
chemistry, we note that the reaction rates are defined over this
temperature range, if not explicitly valid.  However, at all
temperatures, the rates describing the dominant processes are valid.
It is also extremely important to remember
that all \texttt{Grackle} calculations are based on the assumption that the
medium is optically thin.  In practice, the length scale of optical
thickness will become very short as density increases.

\begin{figure*}
  \centering
  \includegraphics[width=0.84\textwidth]{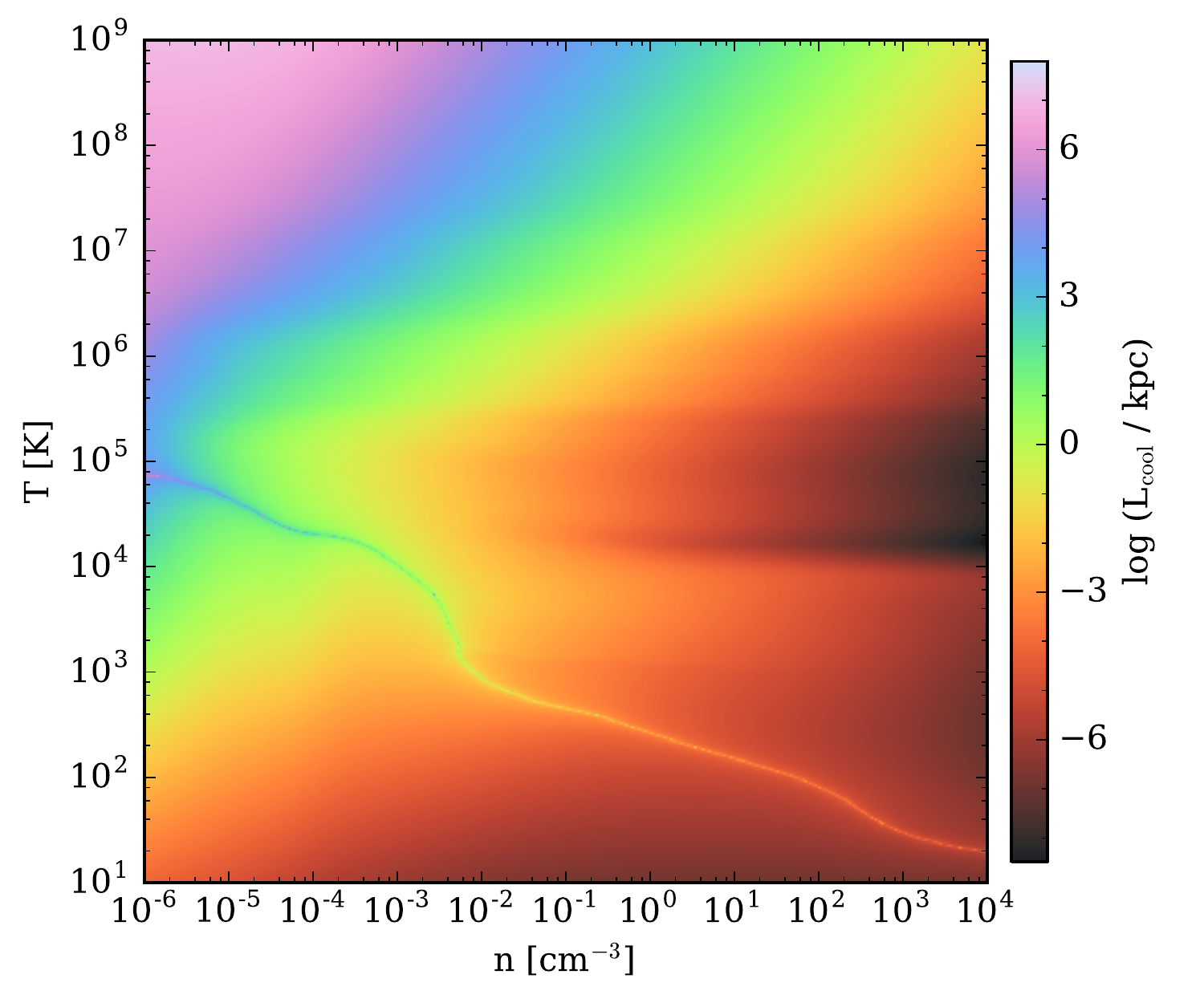}
  \caption{
    The cooling length, defined as the product of the sound speed and
    the cooling time, as a function of number density and temperature
    for a gas with Solar metallicity exposed to a radiation field
    defined by the model of \citet{2012ApJ...746..125H} at z = 0.  The
    narrow line extending from the middle, left to the bottom, right
    represents the temperature where heating and cooling are
    balanced.  Above this line, the gas is being cooled while below
    the line it is being heated.
  } \label{fig:cooling-length}
\end{figure*}

The addition of radiative cooling to a simulation creates another
relevant length scale which must be kept in mind.  The cooling length,
defined as the product of the local sound speed and the cooling time,
sets the approximate size of objects as they cool and condense
\citep{2009A&A...508..725I}.  The cooling length is inversely
proportional to density, effectively setting a density limit for
any given spatial resolution.  When this scale becomes unresolved,
radiative losses will be overpredicted, leading to unphysically high
densities and further exacerbating the resolution problem in a runaway
cycle.  This effect is likely related to the over-cooling problem that
has classically plagued cosmological simulations
\citep[e.g.][]{1996ApJS..105...19K, 2001MNRAS.326.1228B}.
In Figure \ref{fig:cooling-length}, we show an estimate of the
cooling length for the scenario of a gas at Solar metallicity in a
\citet{2012ApJ...746..125H} radiation background at z = 0, noting how
quickly the cooling length drops below 1 kpc, and even 1 pc, for
densities relevant to galaxy formation simulations.  This
length scale should be taken into consideration when choosing the
density threshold above which sub-grid models are applied.

\subsection{Simulation codes with Grackle}
To date, the following codes are known to have \texttt{Grackle}
implemented:
\begin{itemize}

\item AREPO \citep{2010MNRAS.401..791S}

\item ART-I \citep{1999PhDT........25K, 2002ApJ...571..563K}

\item ART-II \citep{2008ApJ...672...19R}

\item CHANGA \citep{2004NewA....9..137W, 2006MNRAS.373.1074S}

\item Cosmos++ \citep{2003ApJS..147..177A, 2005ApJ...635..723A}

\item Enzo \citep{2014ApJS..211...19B}

\item Gadget-3 \citep{2005MNRAS.364.1105S}

\item GAMER \citep{2010ApJS..186..457S}

\item GASOLINE \citep{2004NewA....9..137W}

\item Gear \citep{2012A&A...538A..82R, 2012ASPC..453..141R}

\item Gizmo \citep{2015MNRAS.450...53H}

\item RAMSES \citep{2002A&A...385..337T}

\item SPHS \citep{2012MNRAS.422.3037R}

\item SWIFT \citep{2013arXiv1309.3783G, 2016arXiv160602738S}
\end{itemize}

\subsection{Future Directions} \label{Future_Directions}

\subsubsection{Including new rates and models in Grackle}
The current code structure is highly integrated. This makes introducing new rates for the 
chemical network or cooling function a rather intricate task requiring multiple changes throughout the code. 
Apart from the fact that this is more time consuming it is also much more error prone. In a future release of the 
code the modularity of the code will be greatly increased. There will be a function to populate the species 
rate coefficients and a function to populate the cooling rate coefficients. Separate template files can then be 
updated by a developer wishing to use their own rates. This file can then be included in the build and a flag
set to indicate that the new rates should be used in place of the old rates. Furthermore, a similar method will be 
implemented for solving the network. A template network solver will be available which the developer can use to 
implement a new network with a developer-determined number of species. The developer will be responsible for
updating only three files to achieve a solution to their own chemical network.

\subsubsection{Multiple element cooling}

Currently, \texttt{Grackle} only considers a single metallicity field
for the calculation of the cooling due to heavy elements.  However,
more sophisticated feedback models now consider feedback from multiple
sources, like Type Ia and Type II supernova and winds from AGB stars,
each of which produce distinct abundance patterns.  In the future, we
will look to create additional cooling tables that consider varying
abundance patterns.  As an intermediate step before creating cooling
tables for every metal species, as in \citet{2009MNRAS.393...99W}, we
will likely begin with a two-element model distinguishing between type
Ia and II supernovae, such as that of \citet{2013MNRAS.433.3005D}.

\subsection{Summary}

In this paper, we have described an open-source chemistry and
radiative cooling/heating library suitable for use in numerical
astrophysical simulations.  \texttt{Grackle} includes a number of
non-equilibrium chemistry and cooling models involving H, D and He,
including H$_2$ formation and a simple dust model.  In addition, the
library has the ability to compute equilibrium cooling/heating rates
for primordial and metal-enriched gas, with a number of radiative
backgrounds.  The sophistication of the primordial chemistry network
makes \texttt{Grackle} ideally suited for detailed studies of the
chemistry of metal-free gas.  Although \texttt{Grackle} does not
explicitly follow chemical reactions for elements heavier than He,
the tabulated metal cooling rates allow the code to be employed in all
situations where only the total cooling rate is needed.
The library has an API suitable for calling from C, C++,
FORTRAN and Python.  This paper describes the physical processes
included, the implementation of the models, as well as our open
development and testing framework which allows users/developers to add
to the code in a scalable way that is also intended to minimize new
errors.  We describe the optimization and parallelization strategy,
along with performance benchmarks.  \texttt{Grackle} is well-tested and already
used in a substantial number of high-performance numerical simulation
codes.

%%% Local Variables:
%%% mode: latex
%%% TeX-master: "ms"
%%% End:

\section*{Acknowledgements}

We are grateful to the anonymous referee whose comments helped to
clarify various points in the manuscript.
BDS would like to thank Michael Kuhlen for his work on the code in the
early stages of the project; the organizers of the AGORA project,
Ji-hoon Kim, Joel Primack, and Piero Madau for the original motivation
for the \texttt{Grackle} project; as well as Nick Gnedin, James
Wadsley, and Romeel Dav$\acute{\rm e}$ for providing useful
suggestions for additional functionality.  GLB acknowledges support
from NASA grant NNX15AB20G and NSF grant 1312888. SCOG acknowledges
support from the Deutsche Forschungsgemeinschaft via SFB 881, ``The
Milky Way System'' (sub-projects B1, B2 and B8) and SPP 1573,
``Physics of the Interstellar Medium'' (grant number GL 668/2-1), and
from the European Research Council under the European Community's
Seventh Framework Programme (FP7/2007-2013) via the ERC Advanced Grant
STARLIGHT (project number 339177).  NJG and MJT were supported by NSF
grant ACI-1535651 and by the Gordon and Betty Moore Foundation's
Data-Driven Discovery Initiative through Grant GBMF4651.  JHW is
supported by NSF grants AST-1333360 and AST-1614333 and Hubble theory
grants HST-AR-13895 and HST-AR-14326.  AE is supported by a NSF
Graduate Research Fellowship grant No. DGE-16-44869. We also thank the
NSF for computational resources provided through the XSEDE
program. JAR acknowledges support through the STFC capital grant
ST/L00075X/1 and the Marie Curie Research Fund, grant 699941.
BWO acknowledges support from  NASA through grants
NNX12AC98G, NNX15AP39G, and Hubble Theory Grants HST-AR-13261.01-A and
HST-AR-14315.001-A.  BWO was supported in part by the sabbatical
visitor program at the Michigan Institute for Research in Astrophysics
(MIRA) at the University of Michigan in Ann Arbor, and gratefully
acknowledges their hospitality.
Work by PA was performed in part under the auspices of
the U.S. Department of Energy by Lawrence Livermore National
Laboratiory under Contract DE-AC52-07NA27344.  \texttt{Grackle} was
also made possible by the open-source projects,
HDF\footnote{https://www.hdfgroup.org/},
h5py\footnote{http://www.h5py.org/}, matplotlib
\citep{2005ASPC..347...91B}, and NumPy \citep{numpy}.

%%%%%%%%%%%%%%%%%%%%%%%%%%%%%%%%%%%%%%%%%%%%%%%%%%

%%%%%%%%%%%%%%%%%%%% REFERENCES %%%%%%%%%%%%%%%%%%

\label{lastpage}
\end{document}